\documentclass[]{aa}
\usepackage{graphicx}
\usepackage{txfonts}
\begin{document}
   \title{A possible jet precession in the periodic quasar B0605$-$085}

   \author{
   N.A. Kudryavtseva\inst{1,2,3}\fnmsep\thanks{Member of the International Max Planck Research School (IMPRS) for Astronomy and Astrophysics}
          \and
          S. Britzen\inst{1}
          \and
          A. Witzel\inst{1}
          \and
          E. Ros\inst{4,1}
          \and
          M. Karouzos\inst{1, \star}
          \and
          M. F. Aller\inst{5}
          \and
          H. D. Aller\inst{5}
          \and
          H. Ter\"{a}sranta\inst{6}
          \and\\
          A. Eckart\inst{7,1}
          \and
          J.A. Zensus\inst{1,7}
          }

   \offprints{N.A. Kudryavtseva, nkudryav@mpifr-bonn.mpg.de}

   \institute{Max-Planck-Institut f\"ur Radioastronomie, Auf dem H\"ugel 69,
D-53121 Bonn, Germany
          \and Physics Department, University College Cork, Cork, Ireland
          \and Astronomical Institute of St. Petersburg State University, Universitetskiy Prospekt 28,
          Petrodvorets, 198504, St. Petersburg, Russia
          \and Departament d'Astronomia i Astrof\`{\i}sica, Universitat de Val\`encia, E-46100 Burjassot, Val\`encia, Spain
          \and Astronomy Department, University of Michigan, Ann Arbor, MI 48109, USA
          \and Mets\"{a}hovi Radio Observatory, TKK, Helsinki University of Technology,
          Mets\"{a}hovintie 114, FI-02540 Kylm\"{a}l\"{a}, Finland
          \and I. Physikalisches Institut, Universit\"{a}t zu K\"{o}ln, Z\"{u}lpicher Str. 77, D-50937 K\"{o}ln, Germany
             }

   \date{}


  \abstract
   {
The quasar B0605$-$085 (OH~010) shows a hint for probable periodical variability in the radio total flux-density light curves. }
   {
We study the possible periodicity of B0605$-$085 in the total flux-density, spectra and opacity changes in order to compare it with jet
kinematics on parsec scales. }
   {
We have analyzed archival total flux-density variability at ten frequencies (408~MHz, 4.8~GHz, 6.7~GHz, 8~GHz, 10.7~GHz, 14.5~GHz, 22~GHz,
37~GHz, 90~GHz, and 230~GHz) together with the archival high-resolution very long baseline interferometry data at 15~GHz from the MOJAVE
monitoring campaign. Using the Fourier transform and discrete autocorrelation methods we have searched for periods in the total flux-density
light curves. In addition, spectral evolution and changes of the opacity have been analyzed.
   }
   {
We found a period in multi-frequency total flux-density light curves of $7.9\pm0.5$ yrs. Moreover, a quasi-stationary jet component $C1$
follows a prominent helical path on a similar time scale of 8 years. We have also found that the average instantaneous speeds of the jet
components show a clear helical pattern along the jet with a characteristic scale of 3~mas. Taking into account average speeds of jet
components, this scale corresponds to a time scale of about 7.7 years. Jet precession can explain the helical path of the quasi-stationary jet
component $C1$ and the periodical modulation of the total flux-density light curves. We have fitted a precession model to the trajectory of the
jet component $C1$, with a viewing angle $\phi_{0} = 2.6^\circ \pm 2.2^\circ$, aperture angle of the precession cone $\Omega = 23.9^\circ \pm
1.9^\circ$ and fixed precession period (in the observers frame) $P = 7.9$ yrs.

   }
   {}

   \keywords{Galaxies:active - Galaxies:jets - Radio continuum:galaxies - quasars:general - quasars:individual:B0605$-$085 }

   \maketitle
%
\section{Introduction}

The radio source B0605$-$085 ($OH~010$) is a very bright at centimeter and millimeter wavelengths quasar with a redshift of 0.872 (Stickel et
al. 1993). B0605$-$085 was studied in X-ray and optical wavelengths with {\em CHANDRA} and {\em HST} telescopes by Sambruna et al. (2004) and
is one of a few quasars which jets have been detected in the X-rays. Very Long-Baseline Interferometry (VLBI) observations revealed a complex
structure of the radio jet of the quasar, with multiple bends and curves at scales from parsecs to kiloparsecs. High-resolution space VLBI
observations show that the inner jet extends in the south-east direction and at $\sim$0.2 mas it turns to the north-east (Scott et al. 2004).
When observed at centimeter wavelengths the outer part of the jet continues to extend south-east with multiple turns and at kilo-parsec scales
it goes eastwards with a south-east bend at $\sim$3 arcseconds (Cooper et al. 2007). The kinematics of B0605$-$085 jet was studied by
Kellermann et al. (2004) at 2~cm as part of the MOJAVE 2-cm survey. Based on three epochs of observations (1996, 1999 and 2001), two jet
components were found at 1.6 and 3.8 mas distance from the radio core with moderate apparent speeds of 0.10$\pm$0.21c and 0.18$\pm$0.02c,
respectively. The kinematics of B0605$-$085 and its components accelerations were also studied by Lister et al. (2009) and Homan et al. (2009)
as part of the sample of 135 radio-loud active galactic nuclei. However, only average acceleration of jet components have been discussed in
these papers.

The total flux-density radio light curves of B0605$-$085 from University of Michigan Radio Astronomical Observatory (e.g., Aller et al. 1999)
show hints of periodic variability. Several outbursts appeared in the source at regular intervals. Periodic variability of active galactic
nuclei is a fascinating topic which is not well understood at the moment. Only a few active galaxies show evidence for possible periodicity in
radio light curves (e.g., Raiteri et al. 2001, Aller et al. 2003, Kelly et al. 2003, Ciaramella et al. 2004, Villata et al. 2004, Kadler et al.
2006, Kudryavtseva \& Pyatunina 2006, Qian et al. 2007, Villata et al. 2009). Various mechanisms might cause periodic appearance of outbursts
in radio wavelength, such as helical movement of jet components (e.g., Camenzind \& Krockenberger 1992, Villata \& Raiteri 1999, Ostorero et
al. 2004), shock waves propagation (e.g., G\'omez et al. 1997), accretion disc instabilities (e.g., Honma et al. 1991, Lobanov \& Roland 2005),
jet precession (e.g., Stirling et al. 2003, Bach et al. 2006, Britzen et al. 2010), or they even might be indirectly caused by a secondary
black hole rotating around the primary super-massive black hole in the center (e.g., Lehto \& Valtonen 1996, Rieger \& Mannheim 2000). However,
investigation of repeating processes in active galaxies is extremely complicated due to various factors. The first complicity is that a light
curve is usually a mixture of flares which are most likely not caused solely by one effect. Outbursts might be caused by multiple shock waves
propagation, changes of the viewing angle or other conditions in the jet, such as magnetic field or electron density. Looking for periodicity
is also complicated with possible quasi-periodic nature of variability (e.g., OJ~287, Kidger 2000 and references therein) and long-term nature
of periods, when the observed timescales are of the order of tens of years and really long observations are necessary in order to detect such
timescales. In this paper we aim to study the probable periodic nature of the radio total flux-density variability of the quasar B0605$-$085
and to check what might be a possible reason for it, investigating the movement of the jet components with high-resolution VLBI observations
together with an analysis of spectral indexes and opacity of B0605$-$085.

The structure of this paper is organized as follows: in Sect.~2 we describe the total flux-density variability of B0605$-$085 and focus on
periodicity analysis, spectral changes and frequency-dependent time lags of the flares. In Sect.~3 parsec-scale jet kinematics of B0605$-$085
is presented. In Sect.~4 we apply a precession model to the trajectory of B0605$-$085 jet component. In Sect.~5 we discuss and present the main
conclusions and in Sect.~6 we list main results of the paper.

\section{Total flux-density light curves}
\subsection{Observations}
For investigating the long-term radio variability of B0605$-$085 we use University of Michigan Radio Astronomical Observatory (UMRAO)
monitoring data at 4.8~GHz, 8~GHz and 14.5~GHz (Aller et al. 1999) complemented with the archival data from 408~MHz to 230~GHz.
Table~\ref{data} lists observed frequency, time span of observations, and references. All references in the table refer to the published
earlier data. We analyze total flux-density variability of the quasar at ten frequencies obtained at University of Michigan Radio Astronomical
Observatory, Mets\"{a}hovi Radio Astronomical Observatory, Haystack observatory, Bologna interferometer, Algonquin Radio Observatory, and SEST
telescope. The UMRAO data at 4.8~GHz, 8~GHz and 14.5~GHz since March 2001 are unpublished, whereas data from all other telescopes are archival
and have been published earlier.

\begin{figure*}[htb] \centering  
\includegraphics[clip,width=16.0cm]{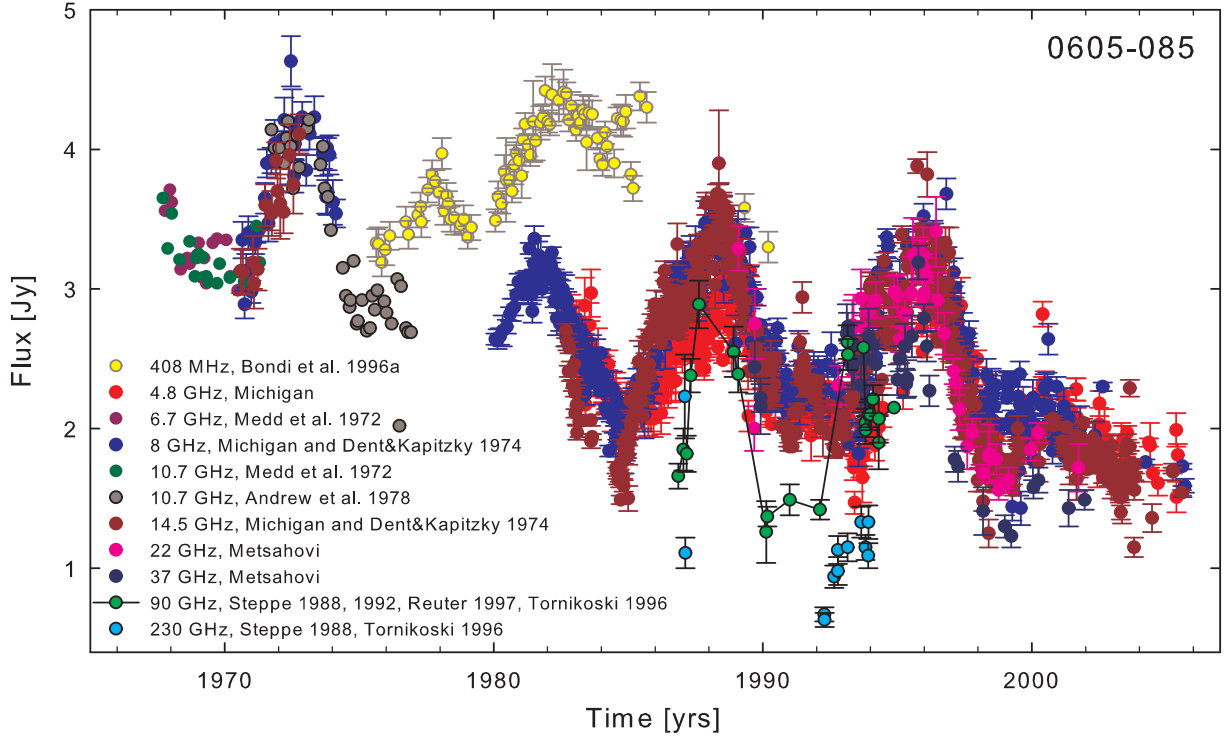}
\caption{Historical total flux-density light curves of B0605$-$085 at ten frequencies from 408~MHz to 230~GHz (see text and Table~\ref{data}).}
\label{0605_hist_lcurve}
\end{figure*}

\begin{figure}[htb] \centering  
\includegraphics[clip,width=7.1cm]{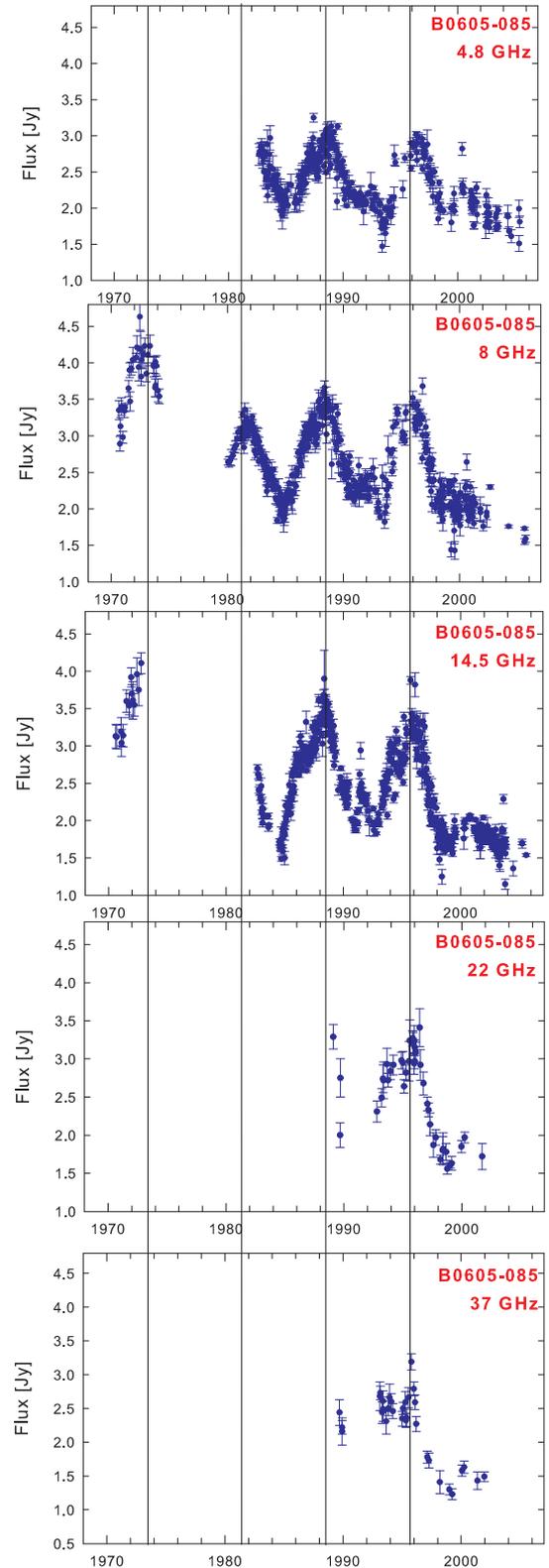}
\caption{Total radio flux-density light curves of B0605$-$085 at five frequencies. The solid lines mark positions of the brightest flares.}
\label{0605_lightcurve}
\end{figure}

The historical light curve is shown in Fig.~\ref{0605_hist_lcurve} and spans almost 40 years of observations. The longest light curve at an
individual frequency covers more than 30 years at 8~GHz (see Table~\ref{data}). The individual light curves at 4.8, 8, 14.5, 22, and 37~GHz are
shown in Fig.~\ref{0605_lightcurve}. The total-flux density variability of B0605$-$085 shows hints for a periodical pattern. Four maxima have
appeared at epochs around 1972.5, 1981.5, 1988.3, and 1995.8 with almost similar time separation. The last three peaks have similar brightness
at centimeter wavelengths, whereas the first one is around 1 Jy brighter. The data at 22~GHz and 37~GHz are less frequent and cover shorter
time ranges of observations, but it is still possible to see that the last peak in 1995-1996 is of similar shape and reaches the maximum at the
same time as at lower frequencies.

\begin{figure}[htb] \centering    
\includegraphics[clip,width=9.0cm]{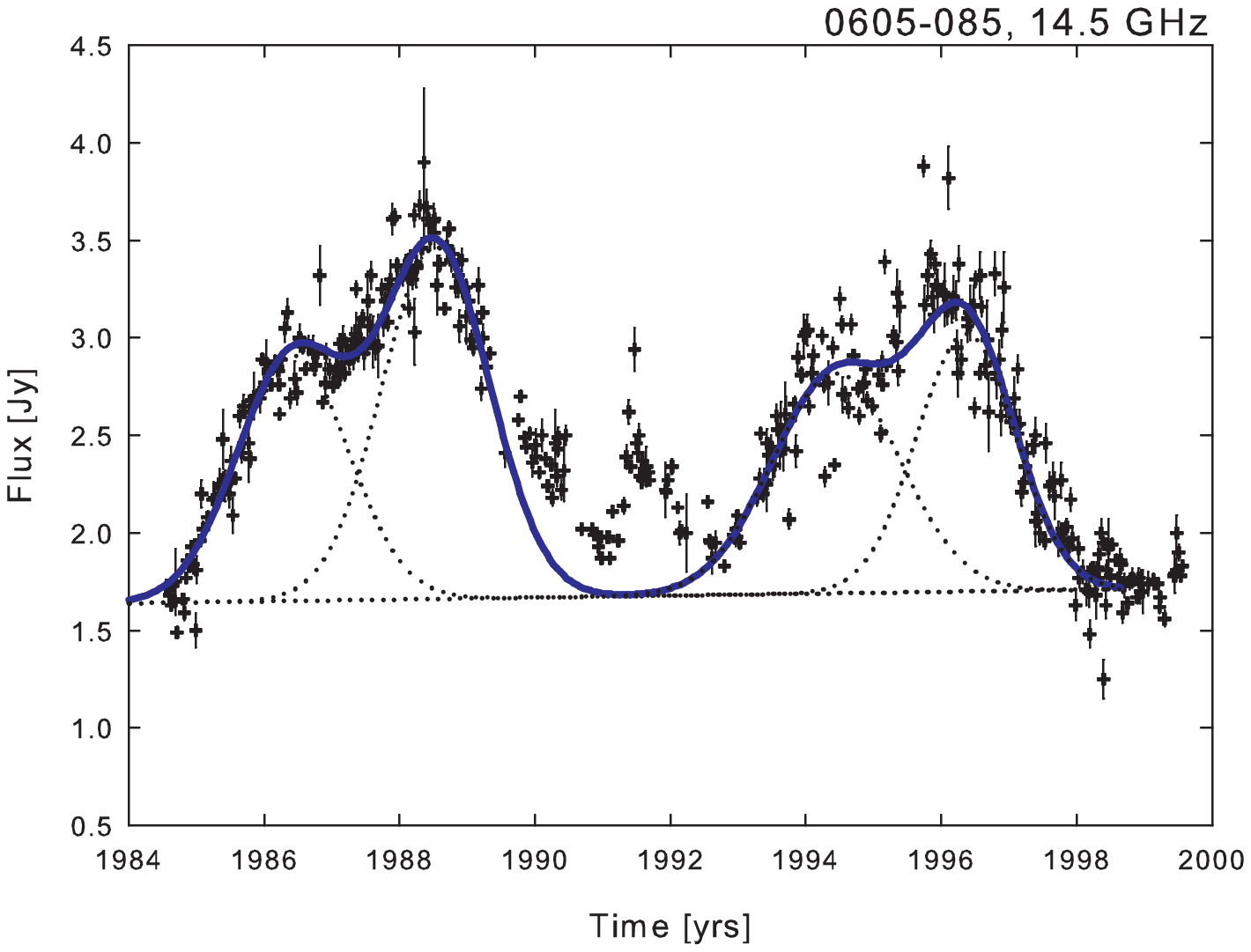}
\caption{Complex sub-structure of flares at 14.5~GHz. The dotted lines show the Gaussian functions fitted to the sub-structure of flares. The
solid line is the sum of fitted Gaussian functions.} \label{0605_double}
\end{figure}

The flares at 14.5~GHz show sub-structure with the dip in the middle of a flare of about 0.5~Jy, which is $\sim$30\% of flare amplitude,
suggesting more complex structure of the outburst or possible double-structure of the flares. Figure~\ref{0605_double} shows a close-up view of
the complex sub-structure of flares at 14.5~GHz. In order to check whether the flares might show the double-peak structure we have fitted the
sub-structure of outbursts at 14.5~GHz with the Gaussian functions (dotted lines in Fig.~\ref{0605_double}). The sum of the Gaussian functions
(solid line) follows the light curve fairly well and the possible double peaks have similar separation, they have time difference of 2.11 years
between 1986.43 and 1988.54 sub-flares and 1.92 years between the 1994.45 and 1996.37 sub-flares.

We searched also for archival data at optical wavelengths. However, due to close proximity of a very bright foreground star of 9th magnitude,
the optical observations of this quasar are difficult and there are not enough data for reconstruction of optical light curves.

\begin{table*}
\caption{List of radio total flux-density observations used in the paper} \label{data} \centering {\small
\begin{tabular}{llll}
\hline \hline
$\nu$ & Time Range & Telescope & Reference \\
\hline

408~MHz  & 1975--1990 & Bologna       & Bondi et al. 1996a \\
4.8~GHz  & 1982--2006 & Michigan      & Aller et al. 1999, Aller et al. 2003 \\
6.7~GHz  & 1967--1971 & Algonquin     & Medd et al. 1972 \\
8~GHz    & 1970--2006 & Haystack, Michigan & Dent \& Kapitzky 1976, Aller et al. 1999, Aller et al. 2003 \\
10.7~GHz & 1967--1971 & Algonquin     & Medd et al. 1972, Andrew et al. 1978 \\
14.5~GHz & 1970--2006 & Haystack, Michigan & Dent \& Kapitzky 1974, Aller et al. 1999, Aller et al. 2003 \\
22~GHz   & 1989--2001 & Mets\"{a}hovi & Ter\"{a}sranta et al. 1998 \\
37~GHz   & 1989--2001 & Mets\"{a}hovi & Ter\"{a}sranta et al. 1998 \\
90~GHz   & 1986--1994 & SEST          & Steppe et al. 1988, Steppe et al. 1992, \\
         &            &               & Tornikoski et al. 1996, Reuter et al. 1997 \\
230~GHz  & 1987--1993 & SEST          & Steppe et al. 1988, Tornikoski et al. 1996 \\

\hline
\end{tabular} }
\end{table*}

\subsection{Periodicity analysis}
In order to check for possible periodical behavior of flares in the B0605$-$085 total flux-density radio light curves we have applied the
Discrete Auto-Correlation Function (DACF) method (Edelson \& Krolik 1988, Hufnagel \& Bregman 1992) and the date-compensated discrete Fourier
transform method (Ferraz-Mello 1981). The discrete auto-correlation function method permits to study the level of auto-correlation in unevenly
sampled data sets avoiding interpolation or addition of artificial data points. The values are combined in pairs $(a_{i}, b_{j})$, for each $0
\leq i,j\leq N$, where N is the number of data points. First, the unbinned discrete correlation function is calculated for each pair
\begin{equation}
\mathrm{UDCF}_{ij} = \frac{(a_{i} - \bar{a})(b_{j} - \bar{b}) }{\sqrt{\sigma_a^2 \sigma_b^2 }},
\end{equation}
where $\bar{a}$, $\bar{b}$ are the mean values of the data series, and $\sigma_{a}$, $\sigma_{b}$ are the corresponding standard deviations.
The discrete correlation function values (DCF) for each time range $\Delta t_{ij} = t_{j} - t_{i}$ are calculated as an average of all UDCF
values, which time interval fall into the range $\tau - \Delta \tau / 2 \leq \Delta t_{ij} \leq \tau + \Delta \tau / 2$, where $\tau$ is the
center of the bin. The higher is the value of $\Delta \tau$ the better is the accuracy and the worse is time resolution of the correlation
curve. In case of the auto-correlation function, the signal is cross-correlated with itself and $a=b$, $\bar{a}=\bar{b}$, and
$\sigma_{a}=\sigma_{b}$. The error of the DACF is calculated as a standard deviation of the DCF value from the group of unbinned UDCF values
\begin{equation}
\sigma(\tau) = \frac{1}{M - 1} \left( \sum [\mathrm{UDCF}_{ij} - \mathrm{DCF}(\tau) ] ^2 \right)^{1/2}.
\end{equation}
The DACF method yields different numbers of UDCF per bin, which can affect the final correlation curve. In order to check the results of the
DACF we also applied the Fisher z-transformed DACF method, which allows to create data bins with equal number of pairs (Alexander 1997).

The Date-Compensated Discrete Fourier Transform (DCDFT) method was created in order to avoid the problem of finiteness of the Fourier transform
and allows us to estimate timescales of variability in unevenly sampled data with better precision. The method is based on power spectrum
estimation fitting sinusoids with various trial frequencies to the data set. In case of a periodic process formed by several waves, the DCDFT
method permits to filter the time series with an already known frequency and find additional harmonics.

\begin{figure*}[htb] \centering    
\hbox{
\includegraphics[clip,width=9.0cm]{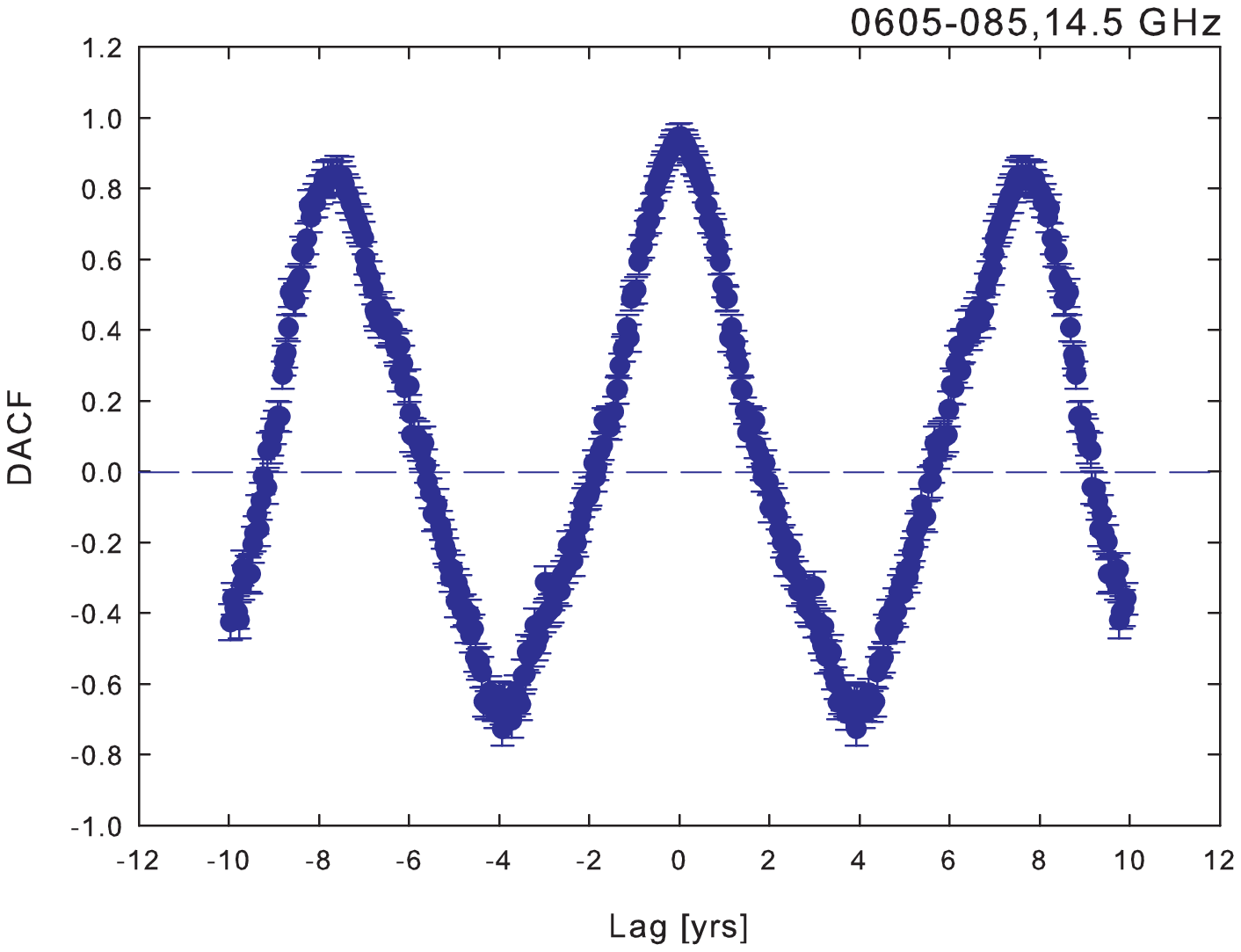}
\includegraphics[clip,width=9.0cm]{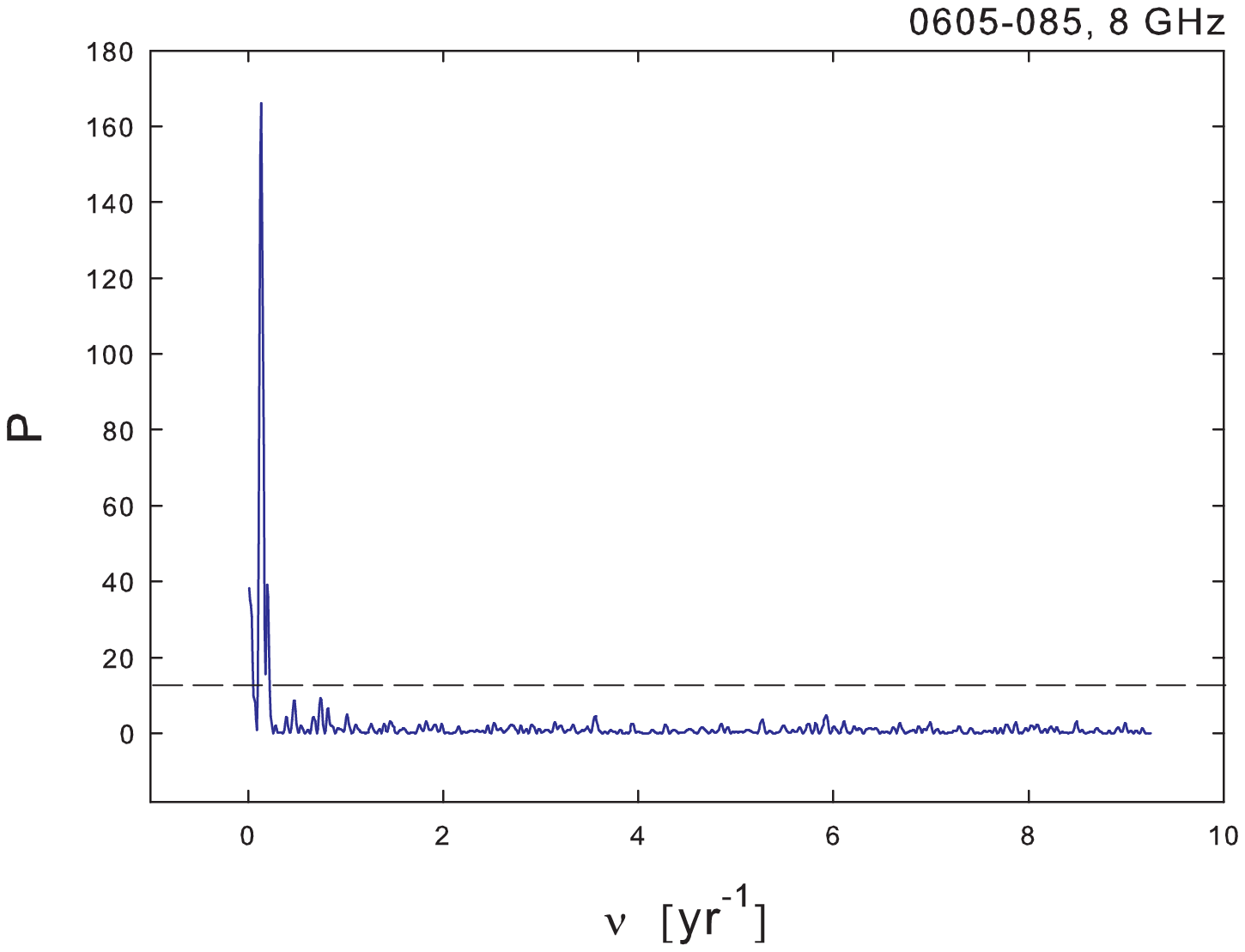}
 }\caption{{\bf
Left:} Discrete auto-correlation function for 14.5 GHz measurements; {\bf Right:} Results of the date-compensated discrete Fourier transform of
the 8~GHz flux-density data. The dashed line marks a 5 \% probability to get a peak by chance.} \label{0605_period}
\end{figure*}

\begin{figure*}[htb] \centering  
\hbox{\includegraphics[clip,width=9.0cm]{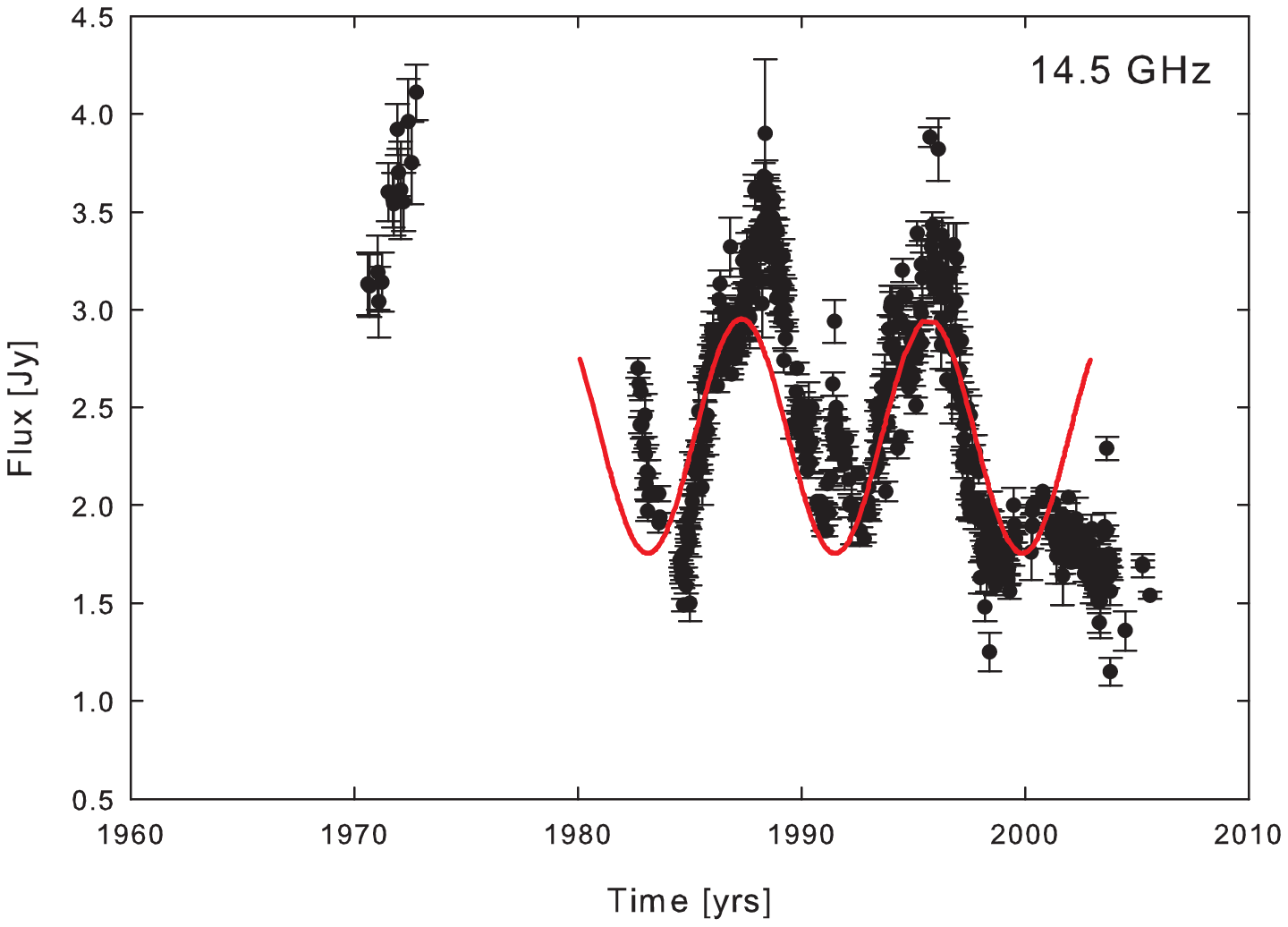}
\includegraphics[clip,width=9.0cm]{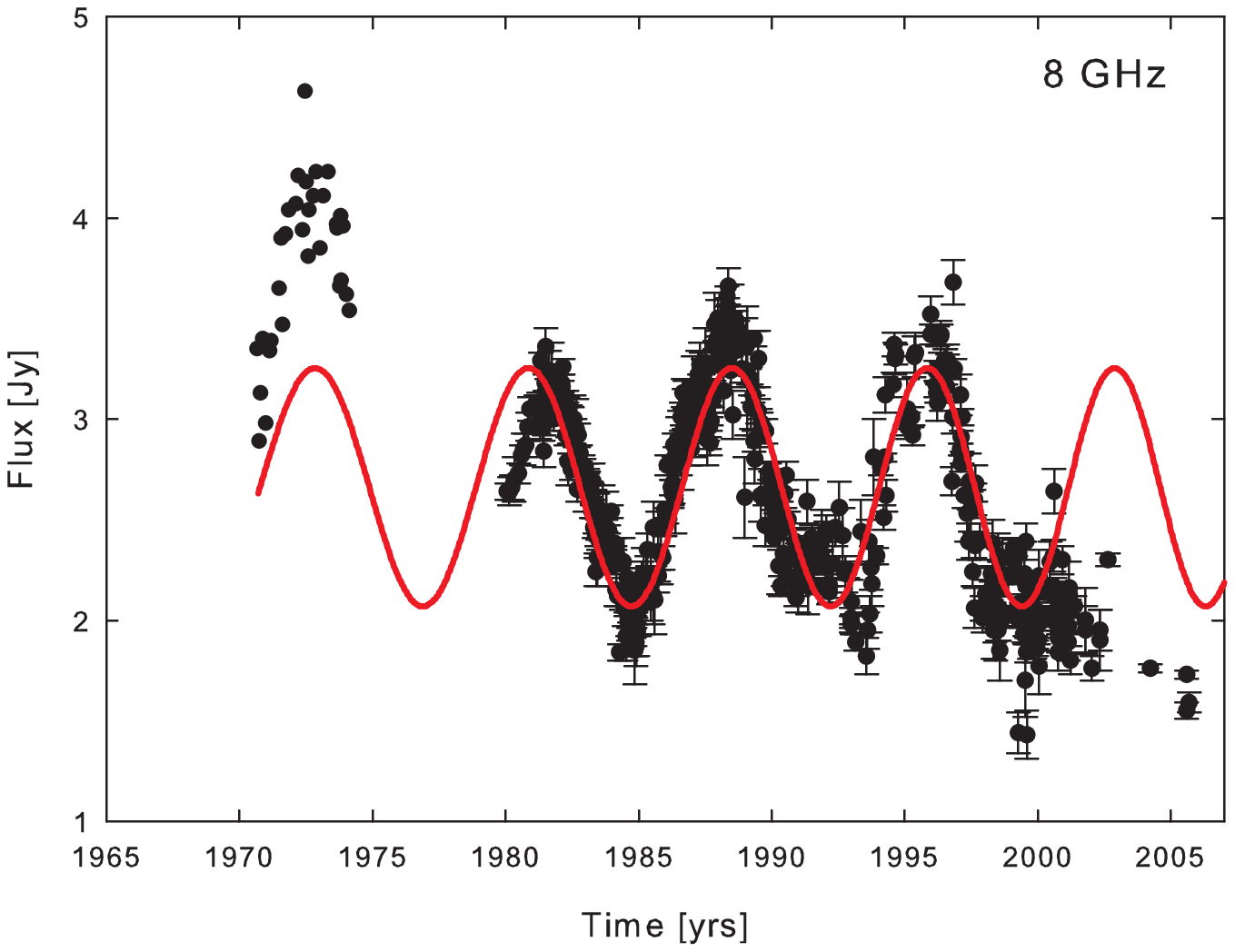}} \caption{
Total flux-density light curves of B0605$-$085 at 14.5~GHz and 8~GHz. The red line shows sinusoids derived from the date-compensated Fourier
transform method with periods of 8.4 and 7.9 years respectively.} \label{0605_dcdft_fit1}
\end{figure*}

\begin{figure}[htb] \centering 
\includegraphics[clip,width=9.0cm]{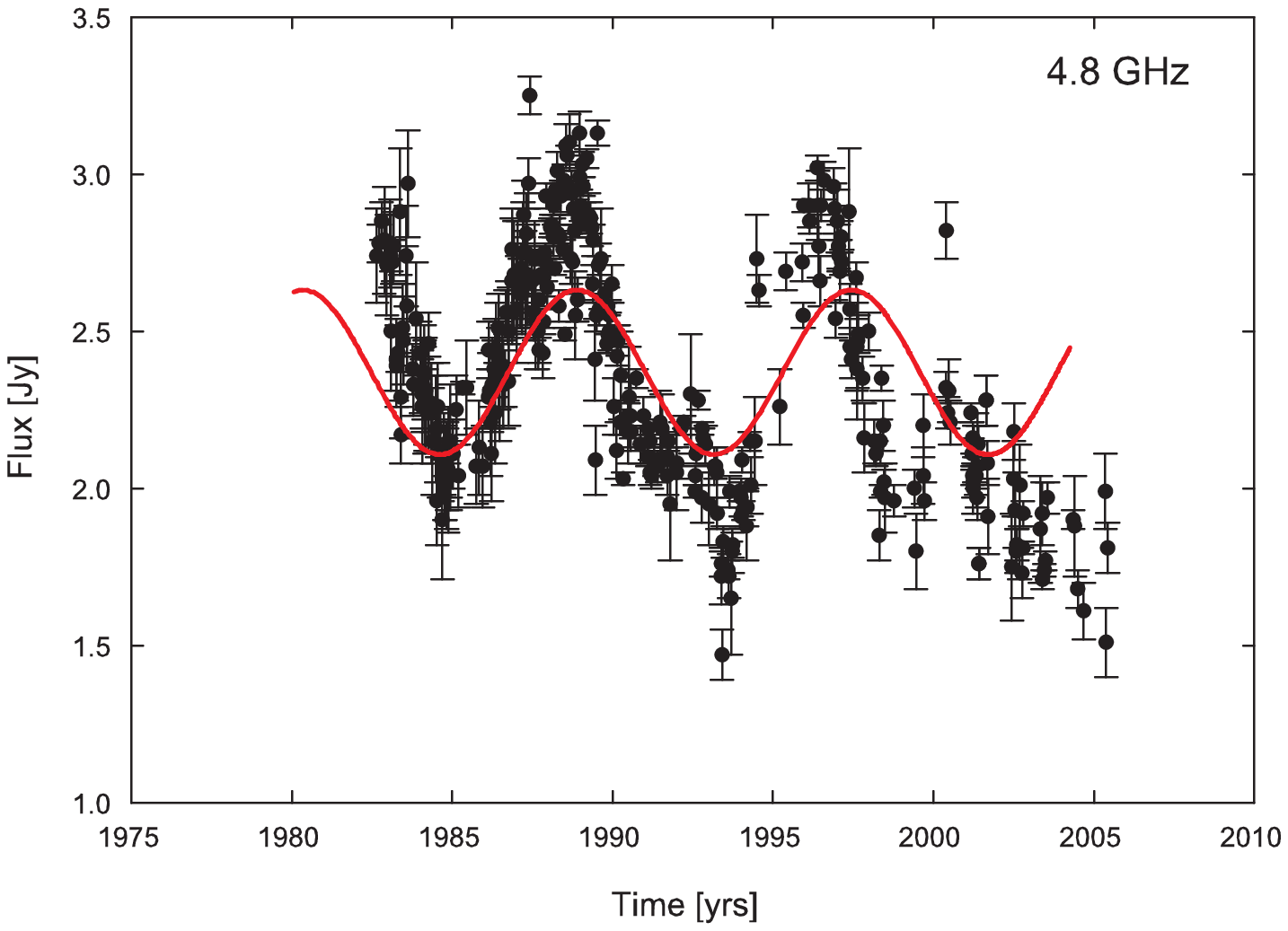}
\caption{Total flux density light curve of B0605$-$085 at 4.8~GHz. The red line shows a sinusoid with a period
of 8.5 years, derived from the date-compensated Fourier transform method} \label{0605_dcdft_fit2}
\end{figure}

We searched for periods in the light curves at 4.8~GHz, 8~GHz, and 14.5~GHz, which span more than 30 years. We have removed a trend before
applying a periodicity analysis. The trend was estimated by fitting a linear regression function into the light curve. The auto-correlation
method reveals periods of 7.7$\pm$0.2 yr at 14.5~GHz, 7.2$\pm$0.1 yr at 8~GHz, and 8.0$\pm$0.2yr at 4.8~GHz with high correlation coefficients
of 0.84, 0.77, and 0.83 respectively. We have estimated the periods by fitting a Gaussian function into the DACF peak. The error bars are
calculated as the error bars from the fit. The Fisher z-transformed method for calculating DACF provides the same results. An example of
calculated discrete auto-correlation function at 14.5~GHz is shown in Fig.~\ref{0605_period} (Left). The plot shows a smooth correlation curve
with a large amplitude of about 0.8, which is an evidence for a strong period. The date-compensated Fourier transform method also shows
presence of a strong periodicity of about 8 years. This method indicates a period of 8.4 years with an amplitude of 0.6~Jy at 14.5~GHz, 7.9
years with amplitude of 0.6~Jy at 8~GHz, and 8.5 years with amplitude of 0.3~Jy at 4.8~GHz. In Fig.~\ref{0605_period} (Right) we plot a
periodogram calculated for the 8~GHz data. The dashed line marks a 5 \% probability to get a periodogram value by chance. The peak
corresponding to the detected period is clearly distinguishable from the noise level and has a high amplitude of about 0.6~Jy. The periodogram
peak reaches 166.01 at the frequency 0.1268 yr$^{-1}$, which corresponds to the period of 7.88 years. The probability to get such peak by
chance is less than $10^{-16}$. The estimated periods are much longer at 4.8~GHz and 14.5~GHz than at 8~GHz, which might be due to different
length of observations at different frequencies. The light curves at 4.8~GHz and 14.5~GHz span about 23 years (if we do not take into account a
few data points in 1970 at 14.5~GHz), whereas the light curve at 8~GHz lasts for 35 years, includes 4 periods and therefore provides more
reliable estimate of the period. The periodicity analysis might show longer periods at 4.8~GHz and 14.5~GHz with the lack of the first peaks in
the light curves. Comparison of the obtained sinusoidal signal, where the wavelength and phase are equal to the measured period, with the
observed light curves is shown in Figs.~\ref{0605_dcdft_fit1} (14.5~GHz and 8~GHz) and \ref{0605_dcdft_fit2} (4.8~GHz). It is clearly seen from
the plots, that these sinusoids follow fairly well the observed flux-density variability for all four flares. The main difference is the
absence of the last peak in 2003--2006 (expected maximum in 2005.9 at 4.8~GHz, 2002.9 at 8~GHz, and 2004.1 at 14.5~GHz), which is predicted by
the periodicity analysis, but is not observed. Both methods give similar results and one can calculate an average period of $7.9\pm0.5$ years
for all three frequencies and for the two methods.

The four main peaks in the total flux-density light curve indicate a periodical behavior of about 8 years. If the period preserves over time,
we would expect a powerful outburst in $\sim$ 2004. However, after year 2000, the flux-density of B0605$-$085 stayed at the same flux level of
about 1.7 Jy at all five frequencies.


\subsection{Frequency-dependent time delays} In case of variability caused solely by jet precession we can
expect that the flares will appear simultaneously at all frequencies. In order to check whether the peaks at different frequencies reach a
maximum at the same time, we calculated frequency-dependent time delays. The Gaussian functions were fitted to the light curves at 4.8~GHz,
8~GHz, 14.5~GHz, 22~GHz, 37~GHz, and 90~GHz as was described in Pyatunina et al. (2006, 2007). The frequency-dependent time delays were
estimated as the time difference between the Gaussian peaks at different frequencies. Due to insufficient data, it was possible to calculate
frequency-dependent time delays only for the outbursts happened in 1988 and 1995-1996. However, from Fig.~\ref{0605_hist_lcurve} it is seen
that the 1973 flare appeared almost simultaneously at 8~GHz and 14.5~GHz. We were not able to get reliable fits for the frequencies 22~GHz and
37~GHz due to sparse observations.

The parameters of the Gaussians, fitted to the light curves, are shown in Table~\ref{0605outbursts}, where frequency, amplitude of a flare,
time of the maximum, width of a flare $\Theta$, and time delay are listed. We calculated frequency-dependent time delays with respect to the
position of the peak at the highest frequency (22~GHz for the 1995-1996 flare and at 90~GHz for the 1988 flare). It is clearly seen that the
$C$ outburst in 1988 appeared almost simultaneously at all four frequencies with the negligible difference in time between individual
frequencies of about 0.08$\pm$0.05 years. The 1995-1996 flare $D$ shows similar properties, the light curves at all frequencies have reached
the maxima almost simultaneously in about 1995.9, except for 4.8~GHz, which has a delay of 0.43$\pm$0.07 years. On the other hand, the data at
4.8~GHz are poorly sampled during the flare rise, which could shift the Gaussian peak. We have also calculated frequency-dependent time lags
using the cross-correlation function. Table~\ref{0605dcftdel} shows delays between various frequencies for $C$ and $D$ flares. The delays for
the $C$ flare are zero within the error bars. The $D$ flare has shown the large time delay of $0.69\pm0.05$ years between 4.8~GHz and 22.0~GHz,
whereas the lags at all other frequencies are much shorter, from $0.08\pm0.02$ to $0.19\pm0.03$ years. These values are consistent with the
lags obtained from the Gaussian fitting, especially that we obtained the larger time delay at 4.8~GHz and lower lags at other frequencies. The
1982 flare is poorly sampled, but it is seen from the plot in Fig.~\ref{0605_hist_lcurve} that it appeared almost simultaneously at 8~GHz and
at low frequency 408~MHz. The 1973 flare was observed only at 8~GHz and 14.5~GHz. There is not enough data for this flare to fit Gaussian
functions, but it is seen from Fig.~\ref{0605_hist_lcurve} that rising of the flare appear simultaneously at two frequencies. Therefore, we can
conclude from the Gaussian fitting and from the visual analysis of the flares, that the 1973, 1988, and 1995-1996 outbursts appeared almost
simultaneously at different frequencies.

The bright outbursts appearing simultaneously at all frequencies can be an evidence for a periodic total flux-density variability caused by the
jet precession. We would expect that if the total flux-density variability is solely due to changes in the Doppler factor, then the peaks of
flares should appear at the same time at various frequencies, since the flux-density $S_{j}$ is changing like $S_{j} = S_{j}' \delta (\phi,
\gamma)^{p+\alpha}$, where $\alpha$ is the spectral index (Lind \& Blandford 1985).

\begin{table}[htb]
\begin{center} \caption{B0605$-$085: Parameters of outbursts} \label{0605outbursts}
\medskip \tiny
\begin{tabular}{lrllr}
 \hline \noalign{\smallskip}
Freq. & Amplitude & $T_{max}$ &$\Theta$ & Time delay \\
{ [GHz]}  &  { [Jy]}         & { [yr]}       &  { [yr]}   &  { [yr]}     \\
\hline \noalign{\smallskip}
C flare & & & & \\
4.8   & $1.36\pm0.01$  & $1988.22\pm0.02$ & $5.61\pm0.02$ & $-0.02\pm0.06$    \\
8.0   & $1.66\pm0.01$  & $1988.29\pm0.01$ & $3.94\pm0.01$ & $0.06\pm0.05$     \\
14.5  & $1.93\pm0.01$  & $1988.30\pm0.01$ & $3.32\pm0.02$ & $0.08\pm0.05$     \\
90.0  & $2.21\pm0.25$  & $1988.22\pm0.04$ & $1.71\pm0.05$ & $0.00\pm0.04$     \\
D flare & & & & \\
4.8   & $1.30\pm0.01$  & $1996.24\pm0.02$ & $4.51\pm0.02$ & $0.43\pm0.07$     \\
8.0   & $1.71\pm0.02$  & $1995.84\pm0.01$ & $3.33\pm0.02$ & $0.03\pm0.07$     \\
14.5  & $1.67\pm0.01$  & $1995.91\pm0.01$ & $3.16\pm0.01$ & $0.09\pm0.07$     \\
22.0  & $1.56\pm0.04$  & $1995.81\pm0.05$ & $2.71\pm0.05$ & $0.00\pm0.05$     \\

\hline
\end{tabular}
\end{center}
\end{table}

\begin{table}[htb]
\begin{center} \caption{B0605$-$085: Time delays estimated with the cross-correlation method} \label{0605dcftdel}
\medskip \tiny
\begin{tabular}{lrr}
 \hline \noalign{\smallskip}
Freq. 1 & Freq.2 & Time delay \\
{ [GHz]}  &  { [GHz]}         & { [yr]}  \\
\hline \noalign{\smallskip}
C flare & &  \\
4.8   & 90.0 & $0.12\pm0.19$ \\
8.0   & 90.0 & $0.34\pm0.08$ \\
14.5  & 90.0 & $0.16\pm0.24$ \\
90.0  & 90.0 & $0.00$        \\
D flare & & \\
4.8   & 22.0 & $0.69\pm0.05$ \\
8.0   & 22.0 & $0.19\pm0.03$ \\
14.5  & 22.0 & $0.08\pm0.02$ \\
22.0  & 22.0 & $0.00$        \\
\hline
\end{tabular}
\end{center}
\end{table}

\subsection{Spectral evolution} 
\begin{figure*}[htb] \centering   
\hbox{\includegraphics[clip,width=9.0cm]{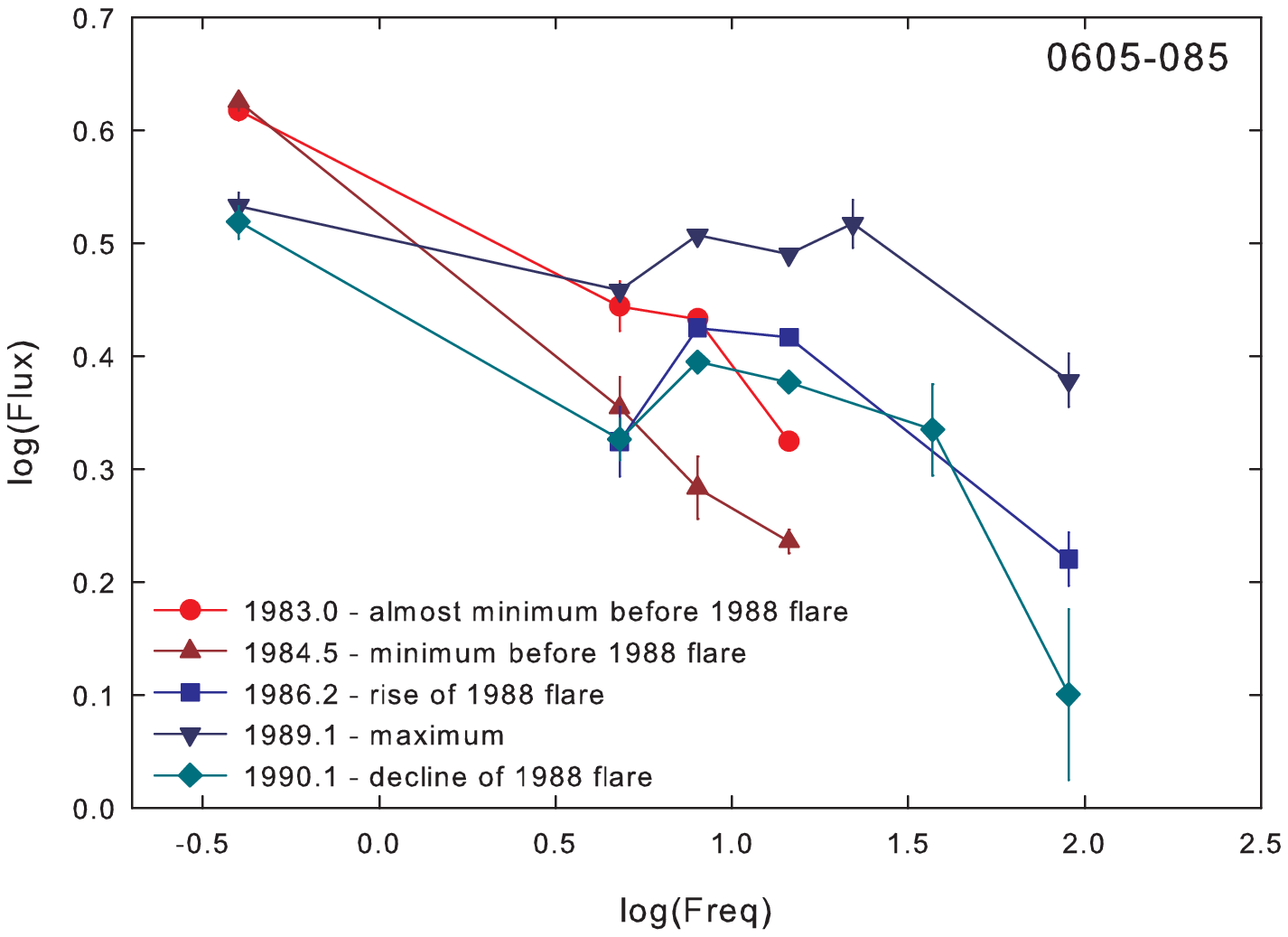} \includegraphics[clip,width=9.0cm]{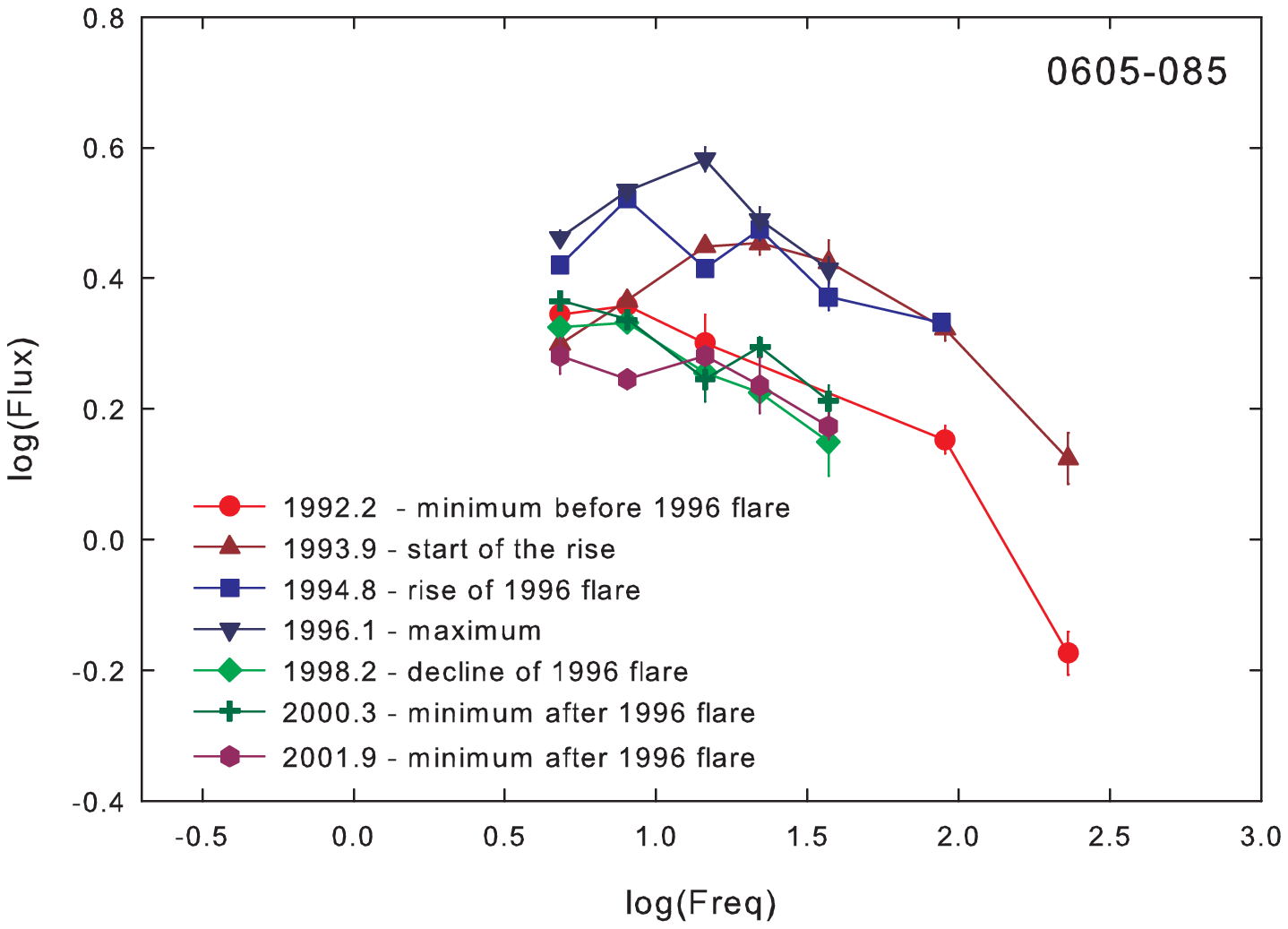}} \caption{Spectral evolution of
B0605$-$085 during the 1988 flare (Left) and 1995-1996 (Right). It is clearly seen how spectral shape changes from steep at the beginning and
at the end of the flares to flat during the maxima.} \label{0605_spectra}
\end{figure*}

In order to check whether all outbursts in the total flux-density light curves
have similar spectral properties, we constructed the quasi-simultaneous spectra for all available frequencies. We have selected and plotted
spectra for various states of variability of B0605$-$085, such as minimum, rising, maximum and falling states. Figure~\ref{0605_spectra} shows
the spectral evolution observed for the 1988 (Left) and 1995-1996 (Right) bright outbursts. The 1988 flare starts from a steep spectrum in 1983
and 1984, that then changes its turnover frequency in 1986. During the maximum of the 1988 flare, the spectrum becomes flat and then steepens
again during the minimum state after the flare. The 1995-1996 flare shows similar spectral evolution, with a steep spectrum at the beginning
and at the end of the flare and a flat spectral shape during the maximum (Fig.~\ref{0605_spectra}).

The light curve of B0605$-$085 consists of four periodical flares in 1973, 1981, 1988, and 1995-1996. The last two flares have similar spectral
evolution and similar brightness. The observations during the 1973 flare are much sparser, but historical radio light curve shows that the
fluxes at 8~GHz and 14.5~GHz are similar, which indicates a flat spectrum. However, it is worth to notice that the second 1981 flare shows a
completely different spectral behavior. One can see from the historical light curve (see Fig.~\ref{0605_hist_lcurve}) that the spectrum of this
flare is steep, since the brightness at 408~MHz significantly exceeds the brightness at centimeter wavelengths. Therefore, we can conclude that
although the periodical flares look almost similar in brightness, they reveal different spectral properties.

\section{Kinematics of B0605$-$085}
\subsection{Observations}
For an investigation of B0605$-$085 jet kinematics we analyze 9 epochs of VLBA observations at 15~GHz from the MOJAVE \& 2cm survey programmes
(http://www.physics.purdue.edu/astro/MOJAVE/). The observations were performed between 1995.6 and 2005.7. The data were fringe-fitted and
calibrated before by the MOJAVE collaboration (Lister \& Homan 2005, Kellermann et al. 2004). Lister et al. (2009) and Homan et al. (2009)
discussed jet components movement and acceleration in this source in a framework of statistical studies of the sample of 135 active galaxies.
In this paper we aim to study motion of B0605$-$085 jet components in more detail, avoiding pre-established initial models during the
model-fitting process.

We performed map cleaning and model fitting of circular Gaussian components using the {\it Difmap} package (Shepherd 1997). We used uniform
weighting scheme for map cleaning. Circular components were chosen in order to avoid extremely elongated components and to ease comparison of
features and their identification. In order to find the optimum set of components and parameters, we fit every data set starting from a
point-like model, adding a Gaussian component after each run of model fitting. The position errors of core separation and position angle were
estimated as $\Delta r = (d \sigma_\mathrm{rms} \sqrt{1 + S_\mathrm{peak} / \sigma_\mathrm{rms}}) / 2 S_\mathrm{peak}$ and $\Delta \theta =
\arctan (\Delta r / r)$, where $\sigma_\mathrm{rms}$ is the residual noise of the map after the subtraction of the model, $d$ is the full width
at half maximum (FWHM) of the component and $S_\mathrm{peak}$ is the peak flux density (Fomalont 1999). However, this formula tends to
underestimate the error if the peak flux density is very high or the width of a component is small. Therefore, uncertainties have also been
estimated by comparison of different model fits ($\pm$1 component) obtained for the same set of data. This permits to check for possible
parameter variations of individual jet components during model-fitting. The final position error was calculated as a maximum value of error
bars obtained by two methods.


\begin{figure}[htb] \centering   
\includegraphics[clip,width=9cm]{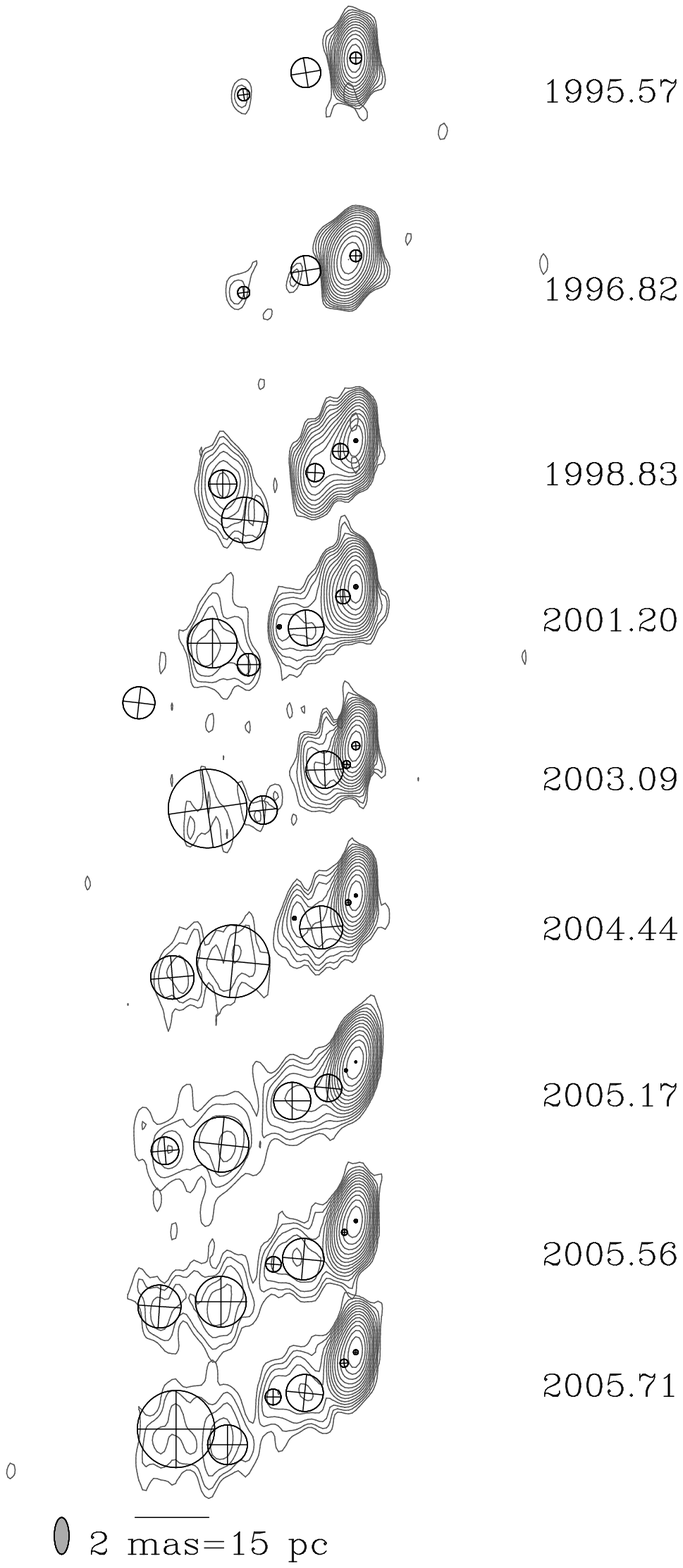} \caption{Contour images of B0605$-$085 and overlapped on them circular
Gaussian components to model the observed visibilities (see Table~\ref{0605modelfits1}). The distance between images on the plot is
proportional to time passed.} \label{0605_maps}
\end{figure}

The final hybrid maps with fitted Gaussian components are shown in Fig.~\ref{0605_maps}. We list all parameters of jet components, such as the
core separation, the position angle, the flux-density, the size, and identification for all epochs in Table~\ref{0605modelfits1}.

\subsection{The component identification}
\begin{figure*}[htb] \centering  
\hbox{ \includegraphics[clip,width=9.0cm]{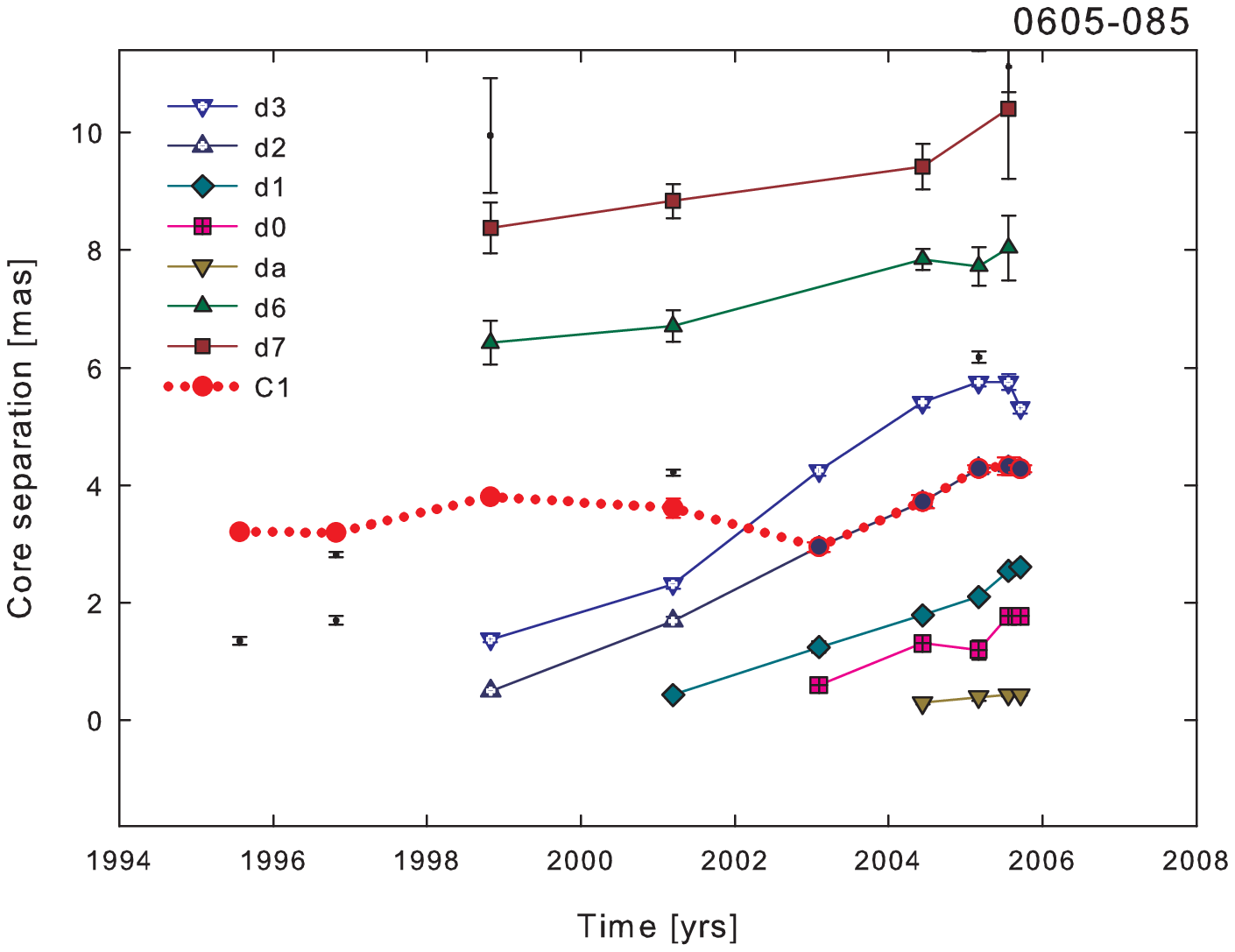}
\includegraphics[clip,width=9.0cm]{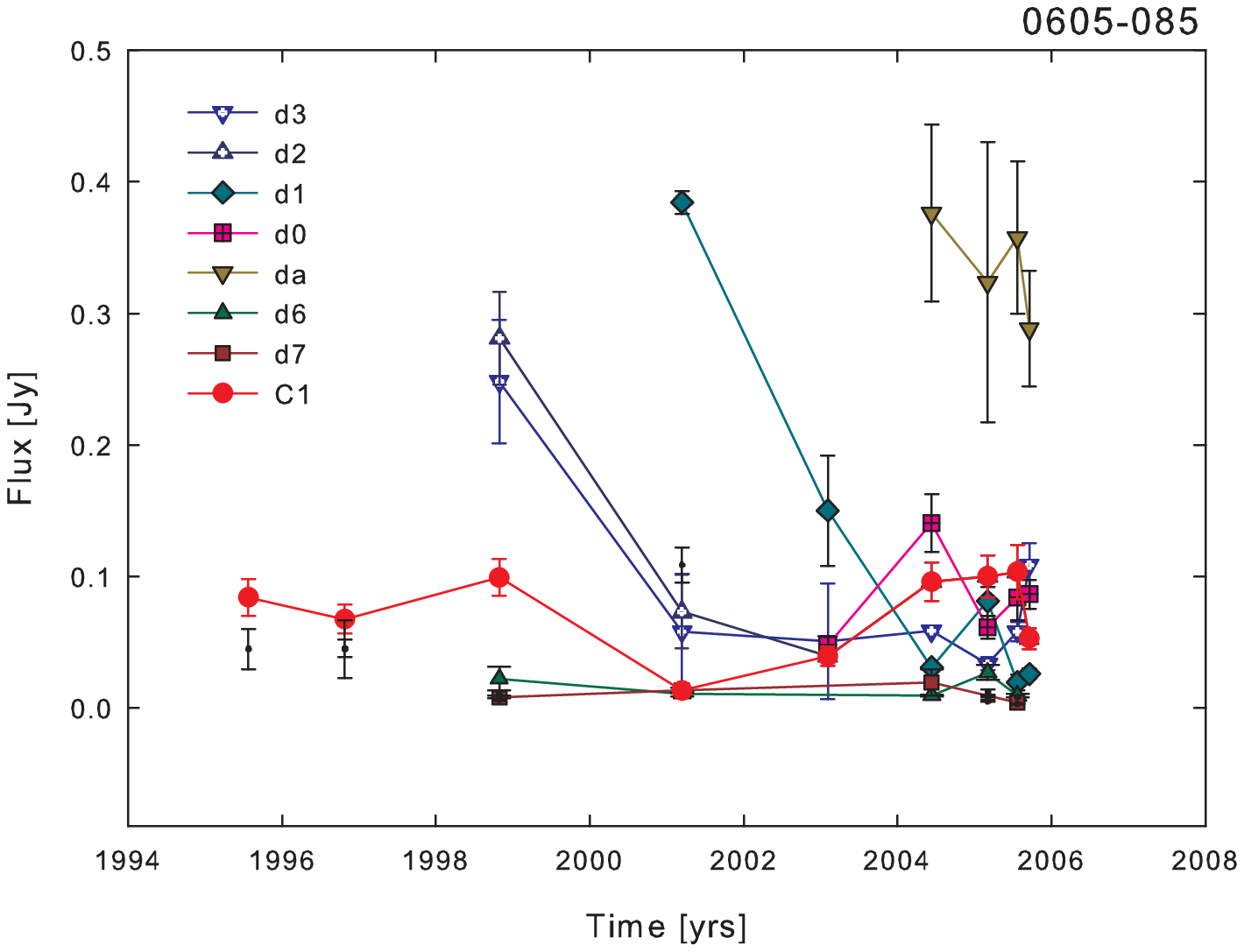}
}\caption{\textbf{Left:} Core separation as function of time at 15 GHz. Individual jet components are marked
by various colors. \textbf{Right:} Time variability of jet components' flux density.} \label{0605_components}
\end{figure*}

The component identification was performed in such way that the core separation, the position angle and the flux of jet features change
smoothly between adjacent epochs. In order to obtain the best identification, the jet components were labelled in different ways and the
identification which gave the smoothest changes in all parameters of jet features, such as mean flux and position, was chosen as final. The
core separations and the flux densities of each individual jet component are shown in Fig.~\ref{0605_components}, left and right
correspondingly. Individual jet features are marked by various colors. Jet feature $C1$ appears quasi-stationary at an average core separation
of 3.7 mas with a flux density of about 0.1 Jy at all epochs. It is marked with a dotted line on the plot. This quasi-stationary feature is
well-detected in the first four epochs, but after 2002 it is blended with the outwards moving jet component $d2$, which was ejected in $\sim$
1998. This feature may be explained with a co-existence of a stationary component and jet features moving outwards. Four outwards moving jet
components ($da$, $d0$, $d1$, $d2$, and $d3$) were ejected during the time of the observations, whereas the components $d6$ and $d7$ are
located at a range of core separations between 6 and 8 mas, showing slow outward movement and have probably been ejected earlier. The jet
components which were ejected at the time of observations are at a level of 0.3 -- 0.4~Jy after ejection and then their flux-density fades down
to 0.1~Jy level or less.

\subsection{Component ejections} We found seven jet components moving outwards with apparent speeds from 0.12
to 0.66 mas/yr, which correspond to speeds of $5.6~c$ -- $31~c$ (see Table~\ref{0605_regression}). The $d3$ component has been ejected in
1996.9 during the 1995-1996 outburst, after the flare has reached its peak. All other components ejected after $d3$ feature, like $d1$, $d0$,
and $da$ show decreasing speeds for almost each successive component. The jet component $d3$ has a speed of $0.66\pm0.01$ mas/yr, the next
feature $d2$ has speed of $0.57\pm0.01$ mas/yr, $d1$ has $0.44\pm0.01$ mas/yr, $d0$ has $0.46\pm0.02$ mas/yr, and the last component $da$ has
$0.12\pm0.03$ mas/yr. A few faint Gaussian components could not be identified between different epochs (marked by black dots in
Fig.~\ref{0605_components}). These components reveal flux-densities on the order of 0.05 Jy, and can either be blobs which were ejected before
1995 and can not be traced because of lack of data, or can be possibly trailing components (e.g., Agudo et al. 2001, Kadler et al. 2008).

The Doppler factor of a jet depends on the Lorentz factor
$\gamma$, the angle between the jet flow direction and the line of sight $\phi$, and the speed of a jet component in the source frame $\beta$:
\begin{equation}
\delta = \frac{1}{\gamma (1 - \beta \cos \phi)}.
\end{equation}
Using velocity of the jet component with the highest speed one can calculate the maximal viewing angle
\begin{equation}
\sin \phi_\mathrm{max} = 2 \beta_\mathrm{app} / (1 + \beta_\mathrm{app}^2)
\end{equation}
and a lower limit for the Lorentz factor
\begin{equation}
\gamma_\mathrm{min} = \sqrt{1 + \beta_\mathrm{app}^2},
\end{equation}
assuming that the highest apparent speed detected defines the lowest possible Lorentz factor of the jet (e.g., Pearson \& Zensus 1986). The
fastest jet component of B0605$-$085 is $d3$, which has a speed of 31c. The corresponding Lorentz factor is $\gamma_\mathrm{min} = 31$ and the
upper limit for the viewing angle is $\phi_\mathrm{max} = 3.7^\circ$, which gives us an estimation for the Doppler factor $\delta_\mathrm{min}
= 12.4$. For smaller viewing angles ($\phi \rightarrow 0$) the Doppler factors approach a value of 62, which yields a range of the Doppler
factors $\delta =$ 12.4 to 62 at viewing angles between $3.7^\circ$ and $0^\circ$.

\begin{table}[htb]
\begin{center}
\tiny \caption{Summary of features' speeds based on the model-fits. Component identification, mean core separation, proper motion of the
component, apparent speed of the component, time of the back-extrapolated component ejection, and the maximum viewing angle are listed.}
\label{0605_regression}
\medskip
\begin{tabular}{lrrlcr}
\hline \noalign{\smallskip} ID & $r_{\rm mean}$ & $\mu_{\rm r}$ &$\beta_{\rm app}$ &$\rm t_{0}$ & $\phi_\mathrm{max}$ \\
 & [mas] & [mas/yr] &[$c$] &[yr] & [deg] \\
\hline \noalign{\smallskip}

$d7$ & $9.26 \pm 0.44$ & $0.20 \pm 0.10$ & $9.4 \pm 4.7 $  &                  & 12.1 \\   
$d6$ & $7.35 \pm 0.32$ & $0.26 \pm 0.06$ & $12.2 \pm 2.8$  &                  & 9.4 \\   
$d3$ & $4.31 \pm 0.67$ & $0.66 \pm 0.01$ & $31.1 \pm 0.5$  & $1996.9 \pm 0.3$ & 3.7 \\
$C1$ & $3.71 \pm 0.17$ & $0.09 \pm 0.01$ & $4.2 \pm 0.5 $  &                  & 26.8 \\
$d2$ & $3.11 \pm 0.56$ & $0.57 \pm 0.01$ & $26.8 \pm 0.5$  & $1997.9 \pm 0.3$ & 4.3 \\
$d1$ & $1.78 \pm 0.34$ & $0.44 \pm 0.01$ & $20.7 \pm 0.5$  & $2000.2 \pm 0.3$ & 5.5 \\
$d0$ & $1.33 \pm 0.22$ & $0.46 \pm 0.02$ & $21.7 \pm 0.9$  & $2001.8 \pm 0.2$ & 5.3 \\
$da$ & $0.39 \pm 0.03$ & $0.12 \pm 0.03$ & $5.6 \pm 1.4 $   & $2001.9 \pm 0.3$ & 20.2 \\

\hline
\end{tabular}
\end{center}
\end{table}

\begin{figure}[htb] \centering  
\includegraphics[clip,width=9.0cm]{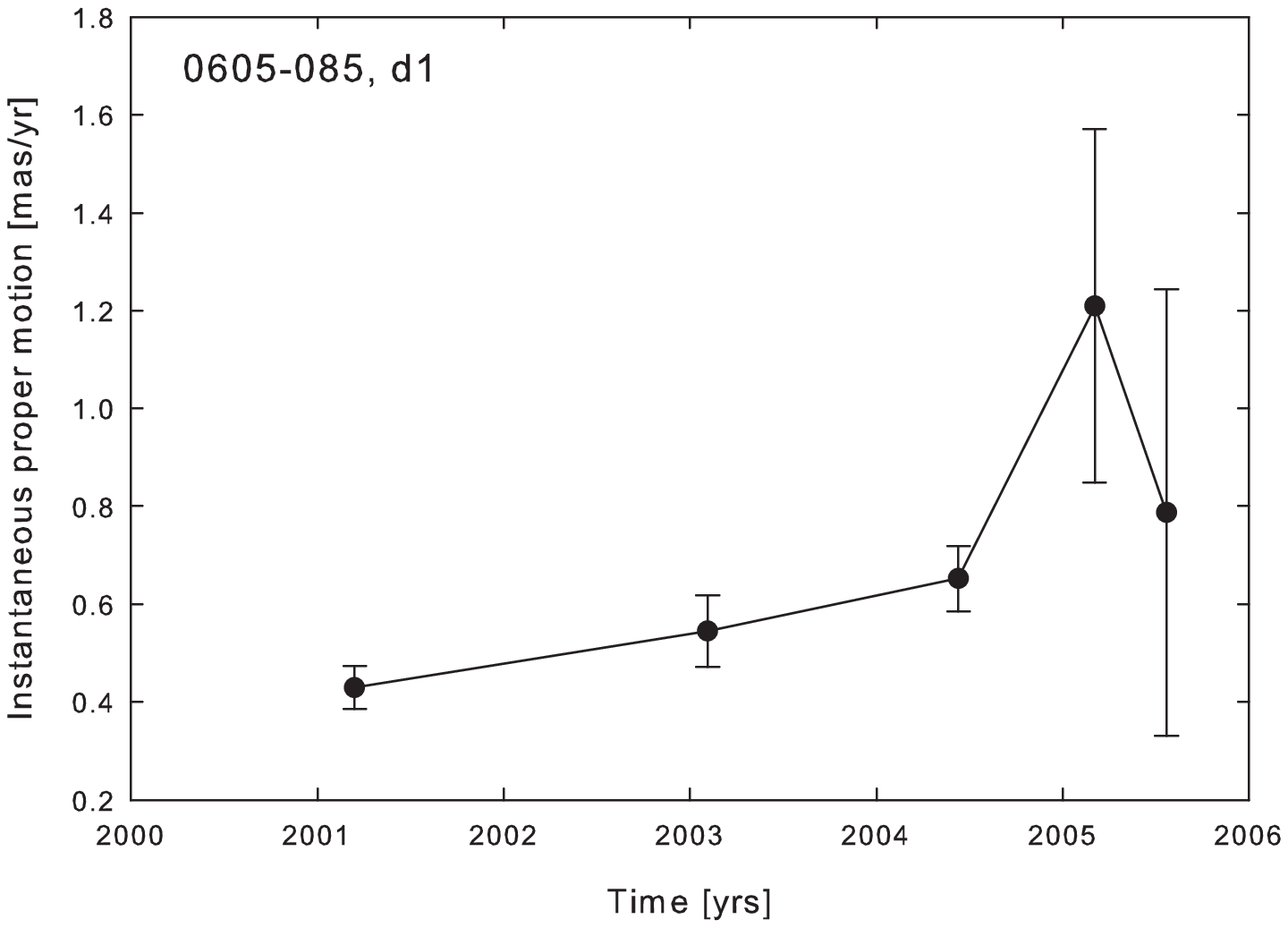}
\caption{Instantaneous speeds of the jet component $d1$, which experiences significant acceleration when moving along the jet.}
\label{0605_d1speed}
\end{figure}

\begin{figure}[htb] \centering  
\includegraphics[clip,width=9.0cm]{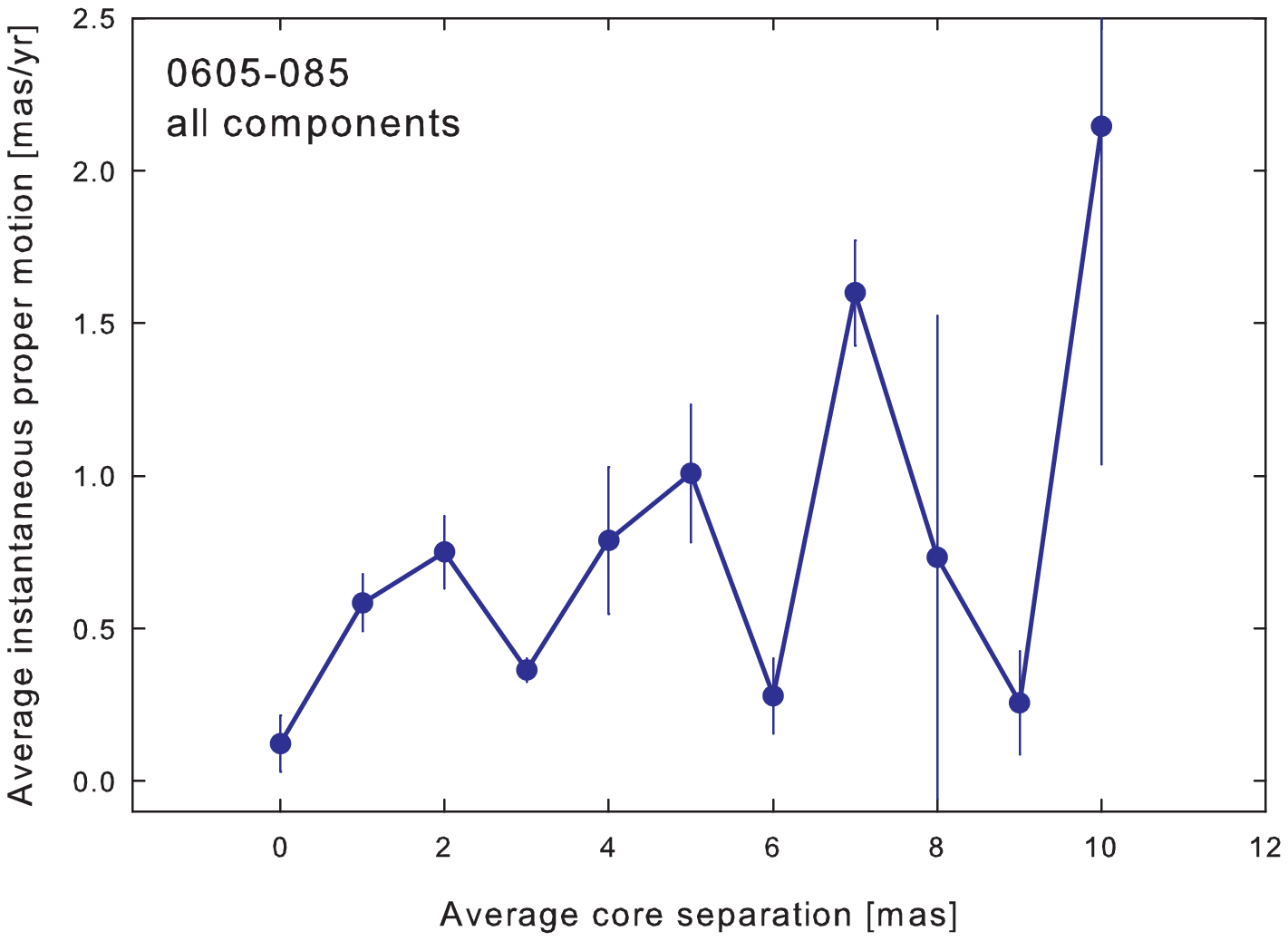}
\caption{Instantaneous speeds of all jet components of B0605$-$085 averaged in 1~mas core separation bins. We interpret that the jet is helical
with a characteristic scale of 3 mas.} \label{0605_instant_speeds}
\end{figure}

The instantaneous speeds of the jet components have been calculated as the core separation difference between two subsequent positions of a jet
component divided by the time elapsed between these two observations. The jet components $d1$, $d3$, and $C1$ show significant changes of
instantaneous speeds which indicates that they experience acceleration and deceleration along the jet (see Fig.~\ref{0605_d1speed} for the jet
component $d1$). Figure~\ref{0605_instant_speeds} shows the averaged instantaneous speeds of all jet components in the 1-mas core separation
bins. We have calculated instantaneous speeds between each data point for each jet component and then have averaged all the instantaneous
speeds for data points located within a certain bin of core separation. The average instantaneous speeds of all jet components accelerate at
about 0.1~mas, 3~mas, 6~mas, and 9~mas distance from the core and decelerate at 2~mas, 5~mas, and 7~mas, which indicates a characteristic scale
of 3 mas for acceleration/deceleration of the jet components. The helical pattern of the speeds might be due to a helical pattern of the jet
and multiple bends on the way of the jet components. It is worth to notice, that the characteristic spatial scale of 3~mas corresponds to a
timescale of about 7.7 years (using the average speed of the jet components 0.39 mas/yr), which is similar to the periodic timescale of
variability in the total flux-density.

\subsection{Quasi-stationary component $C1$}
The quasi-stationary component $C1$ is a well-defined feature from 1995 to 2001 (see Fig.~\ref{0605_components}). However, after 2002 it is
most probably blended with the moving jet component $d2$. Earlier than 2002, when the quasi-stationary component is clearly identified, there
were no moving jet components. That might explain why the component $C1$ was blended only after 2002. As the newly ejected jet feature approach
the position of component $C1$, it becomes more complicated to distinguish between them and it is more likely that we see blending of two
components. Blending and dragging of stationary jet components by moving components was predicted from the magneto-hydrodynamical simulations
by Gomez et al. (1997) and Aloy et al. (2003) and was observed in a large number of sources (e.g., 3C~279 Wehrle et al. (2001), S5~1803+784
Britzen et al. (2005), 0735+178 Gabuzda et al. (1994)).

\begin{figure}[htb] \centering  
\includegraphics[clip,width=9.0cm]{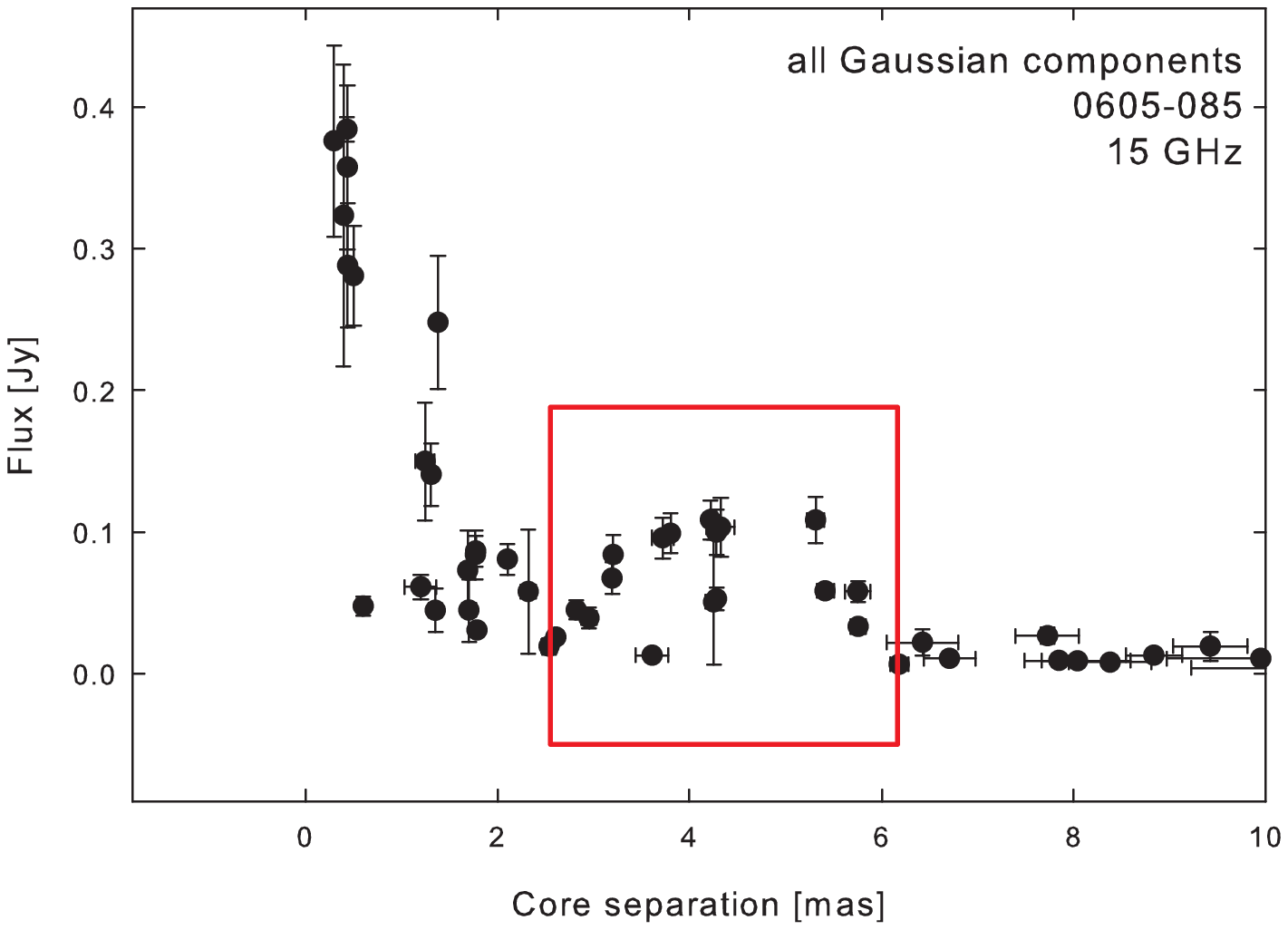}
\caption{Flux-density of all Gaussian components found in the jet of B0605$-$085 at 15~GHz. The core was excluded from the plot for smaller
range of flux densities. It is clearly seen that all components become significantly brighter when they are at the range of the core
separations from 2.5 to 6.2 mas, which can be explained by the jet bending.} \label{0605_flux_r}
\end{figure}

The quasi-stationarity of jet component $C1$ can be due to various physical mechanisms. It can be caused by jet bending or by stationary
conical shock waves appearing due to interaction with the ambient medium (e.g., Alberdi 2000, Gomez et al. 1997). If the stationary feature in
the jet is caused by jet bending towards us, then we can expect a significant flux-density increase in other jet components in the area close
to the position of the stationary feature due to relativistic boosting. When we plot the flux-density versus core separation for all detected
Gaussian components (excluding the core) in Fig.~\ref{0605_flux_r} we see that at core separations between 2.5 and 6.2 mas there is a
significant increase in flux-density values. This increase is independent of the component identification and might be explained by a jet
bending in the area around 4 mas. Moreover, this jet bending can explain the existence of the quasi-stationary jet feature $C1$, which is
located at similar core separation of about 3.7 mas.

High-resolution VLBI maps of the quasar B0605$-$085 were discussed in several papers (e.g., Padrielli et al. (1986), (1991), Bondi et al.
(1996b)). A stationary jet component observed at 2 and 5~GHz in 1997 at a core separation similar to $C1$ was reported before by Fey\&Charlot
(2000). They found the jet components at a core separation of 3.0 mas and $PA= 100^\circ$, which is similar to the position of the $C1$
component. This is an additional evidence for stationarity of the jet component $C1$.

\begin{figure*}[htb]   
\hbox{
\includegraphics[clip,width=6.0cm]{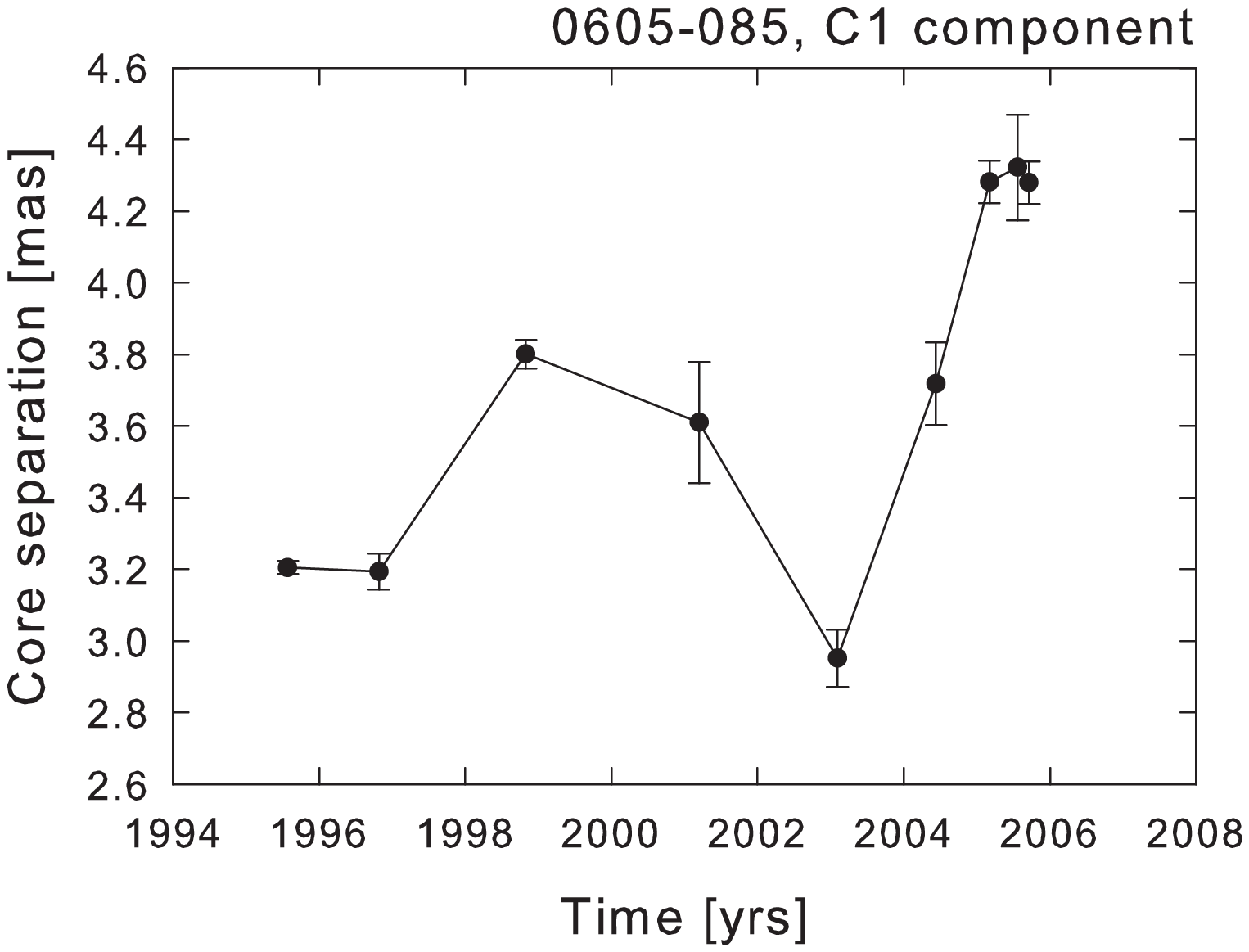}
\includegraphics[clip,width=6.0cm]{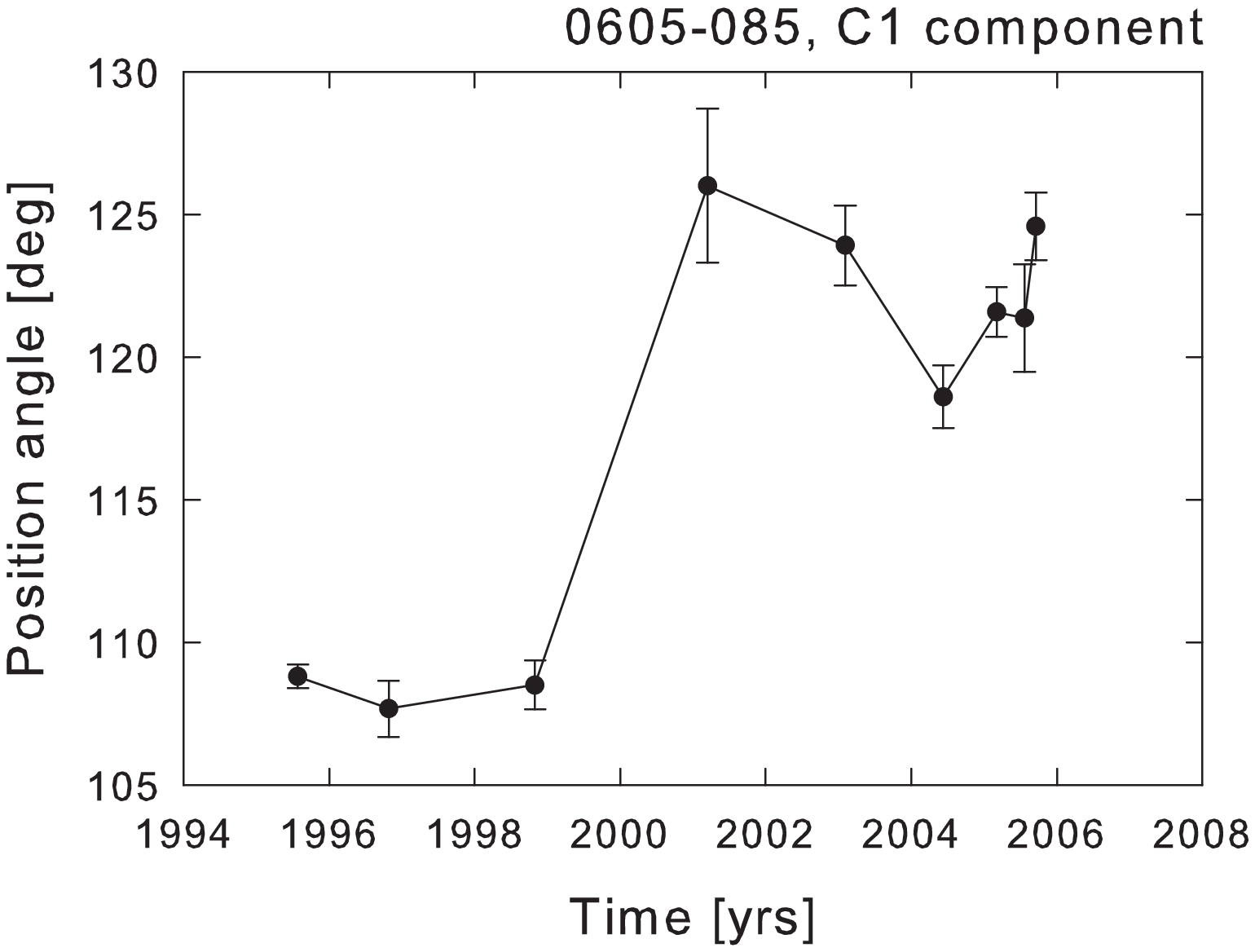}
\includegraphics[clip,width=6.0cm]{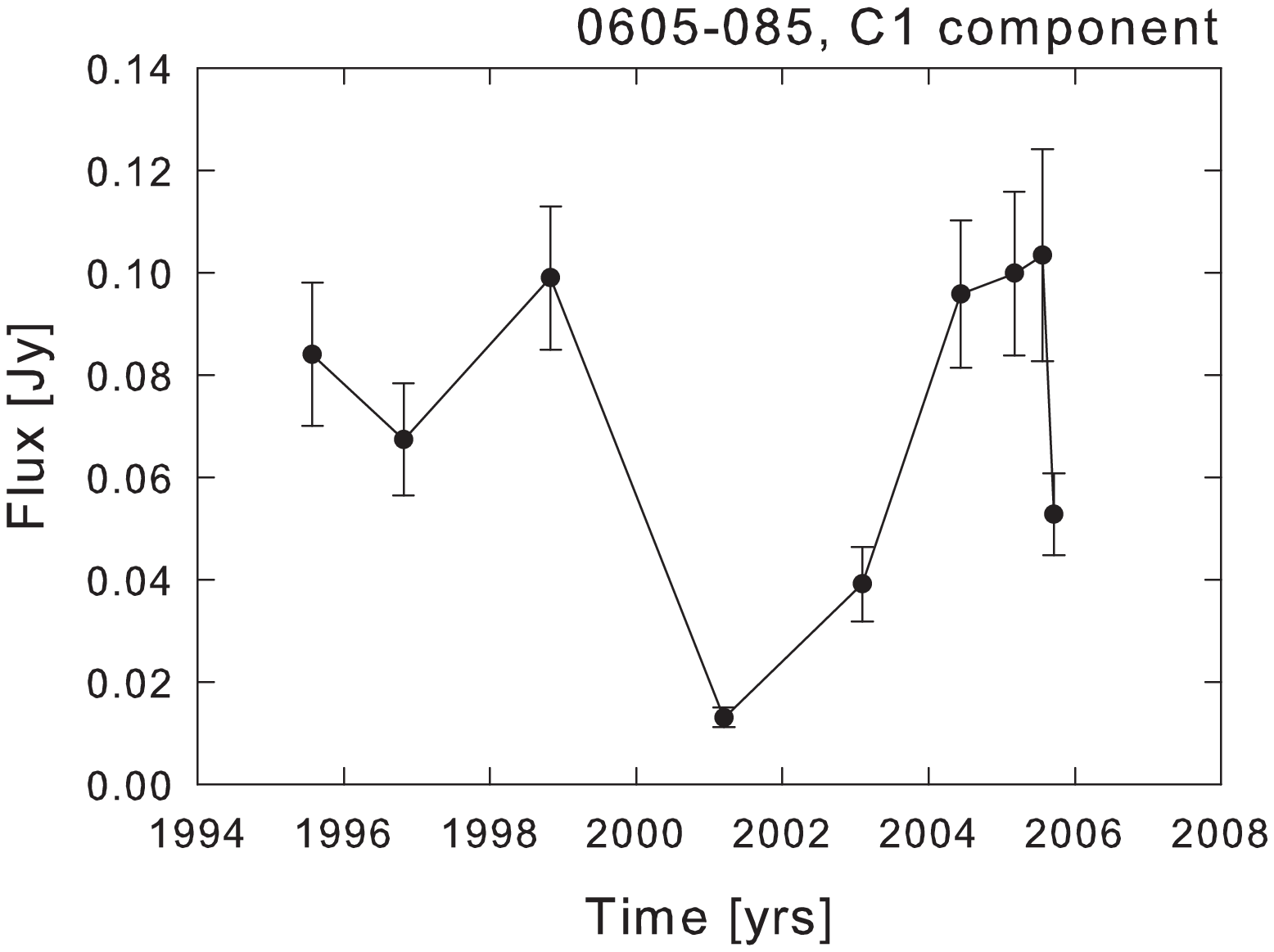}
}\caption{Time evolution of core separation (Left), position angle (Middle) and flux-density (Right) of the quasi-stationary component $C1$.}
\label{0605_c1}
\end{figure*}

\begin{figure*}[htb] \centering  
\hbox{
\includegraphics[clip,width=9.0cm]{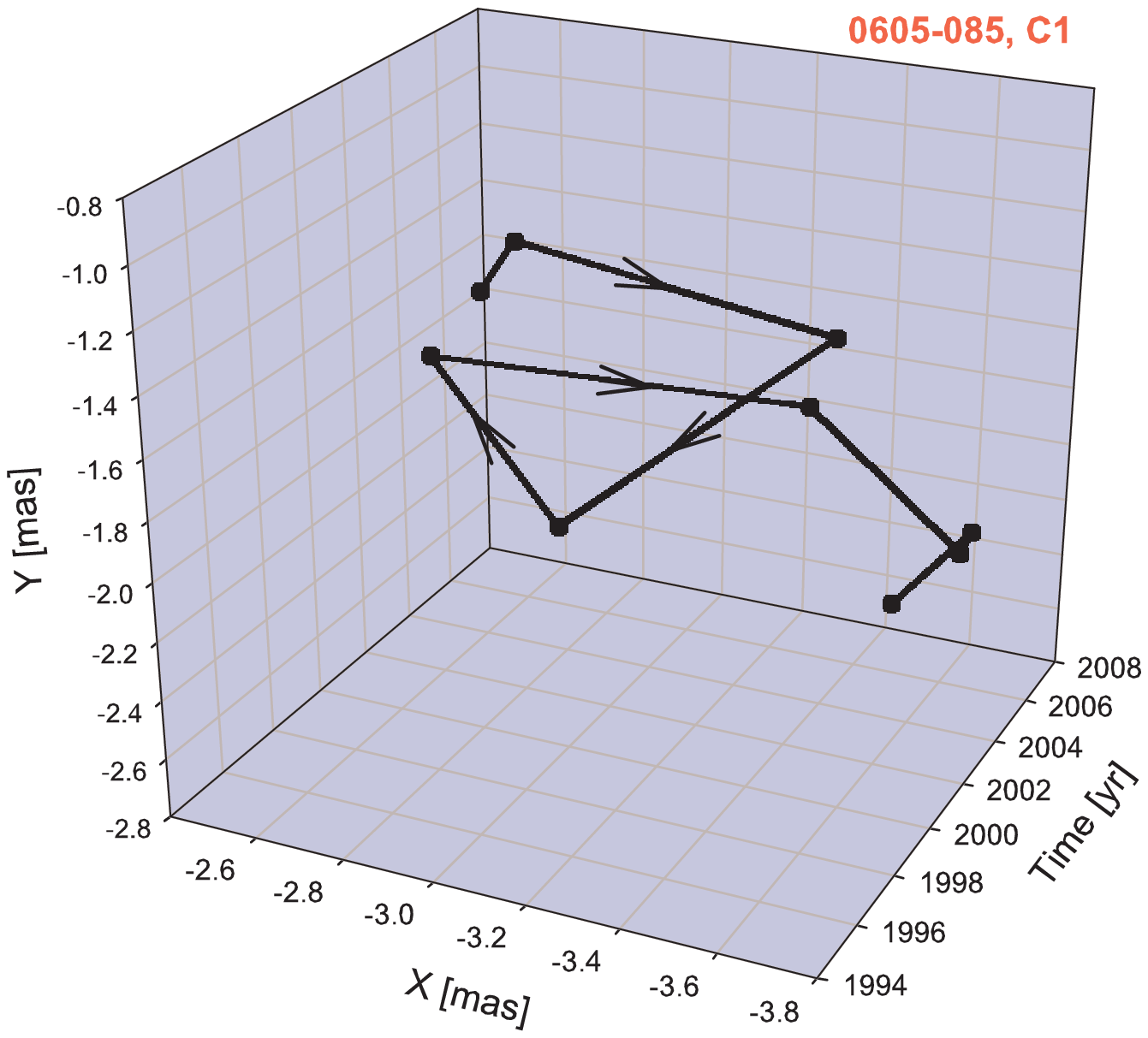}
\includegraphics[clip,width=9.0cm]{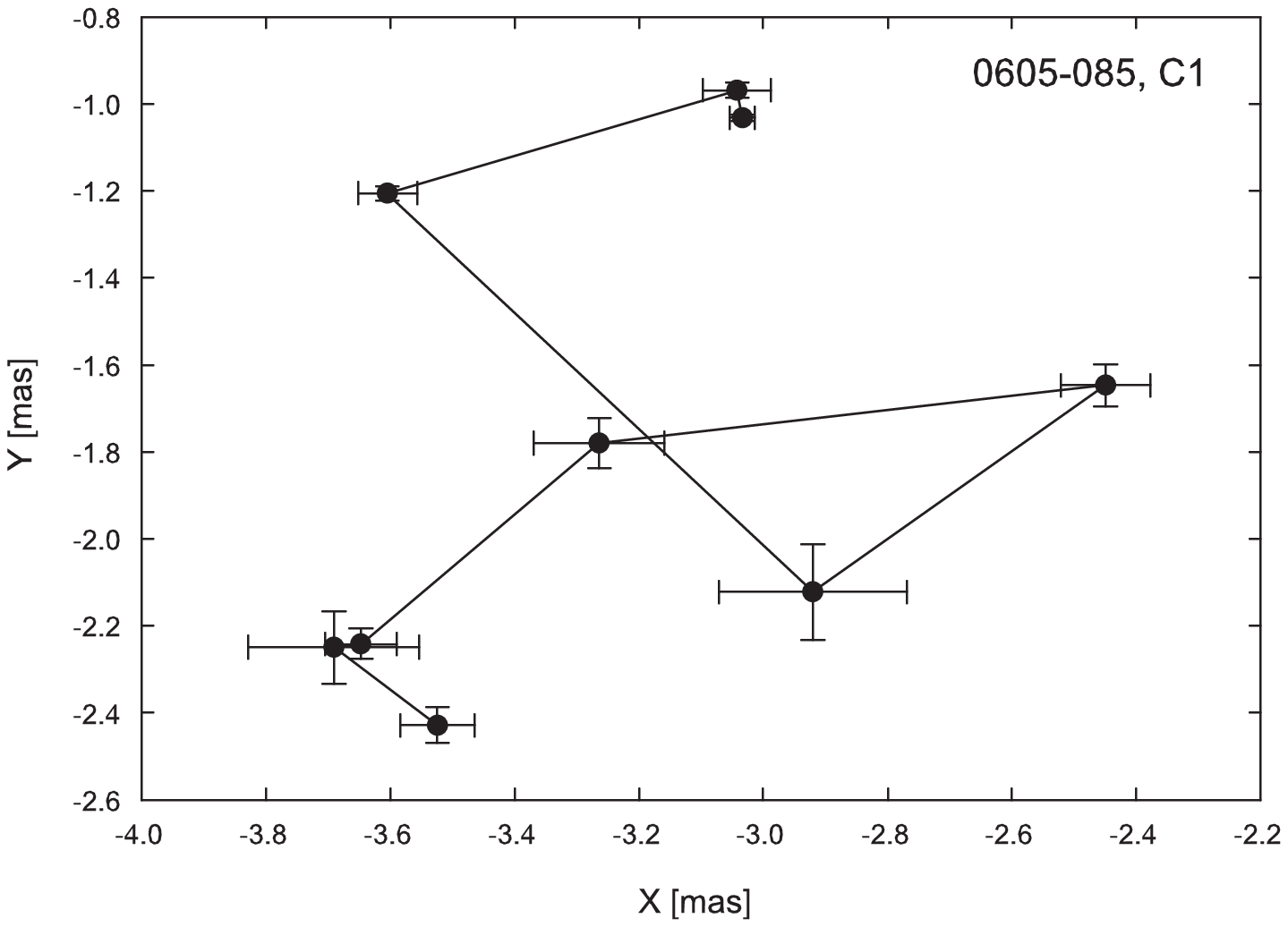}
}\caption{Trajectory of the quasi-stationary component $C1$ in rectangular coordinates in three-dimensional space with the third axis showing
the time (Left) and in two dimensions (Right). The trajectory of the component seems to follow a helical path.} \label{C1_3D}
\end{figure*}

The quasi-stationary jet component $C1$ is located at an average core separation of 3.7 mas. However, the position of this component changes
with time from 2.95 to 4.32 mas. The position angle and the flux-density of $C1$ are also variable. Figure~\ref{0605_c1} shows these changes
over time in core separation, position angle and flux-density. The position angle and the flux-density of the component are anti-correlated.
The position angle reaches its maximum value of 125 degrees in 2001, when the flux is at the minimum value of 0.013$\pm$0.002 Jy. There is also
a possible connection between the core separation and the flux-density of the component $C1$ -- both of them have minima in about 2001--2003
and then experience a rise with a maximum in 2005. In rectangular coordinates, the trajectory of the quasi-stationary jet component $C1$
follows a clear helical path. Figure~\ref{C1_3D} shows a trajectory of $C1$ in the plane of the sky. It is clearly seen from the plot that the
quasi-stationary component follows a spiral trajectory. Moreover, the characteristic time scale of a helix turn is about 8 years. This time
scale is very close to the value of the period of the total flux-density variability discussed earlier. This might be evidence for a possible
connection between the helical movement of $C1$ and the total flux-density periodicity. The jet might precess and therefore due to variability
of the Doppler boosting we will see periodic flares in the light curves as well as changes of position of a bend in the jet.

\subsection{Comparison of trajectories} 
\begin{figure}[tb] \centering  
\includegraphics[clip,width=9.0cm]{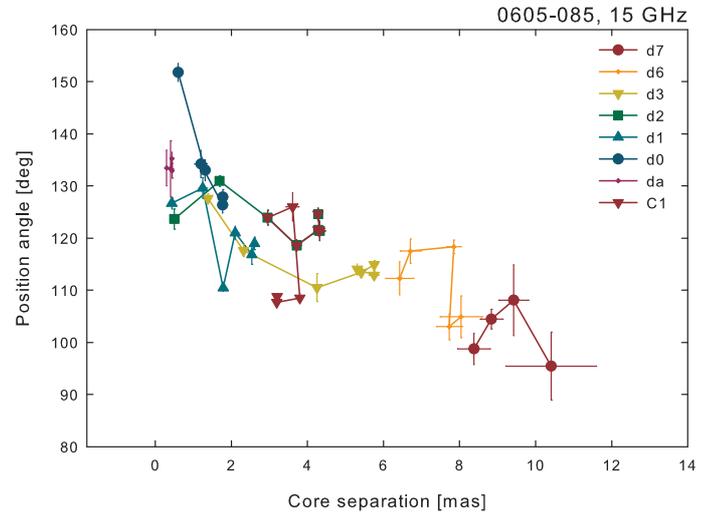}
\caption{Position angle changes of the jet components $da$--$d7$, including the quasi-stationary component $C1$ at 15~GHz.} \label{0605_pa}
\end{figure}

We plot the position angles of all features $da$--$d7$, including the
quasi-stationary component C1 versus the core separation in Fig.~\ref{0605_pa}. The trajectories of the components $d1$, $d2$, $d6$, $d7$, and
$C1$ follow significantly curved trajectories, whereas the trajectories of the components $d0$ and $d3$ follow a linear path. This can be an
evidence for a possible co-existence of two types of jet component trajectories, similar to the source 3C~345 (Klare et al. 2005). The spread
of position angles of all components is changing along the jet. In the inner part, within the first 1 mas, the position angles are in the range
of 123--150 degrees, in the middle part of the jet between 1 -- 5 mas, the position angles range from $110^\circ$ to $125^\circ$, whereas in
the outer part from 5 mas to 10 mas, from $95^\circ$ to $115^\circ$.

\begin{table*}[htb]
\begin{center}
\caption{\tiny Model-fit results for B0605$-$085 at 15 GHz. We list the epoch of observation, the jet component identification, the
flux-density, the radial distance of the component center from the center of the map,  the position angle of the center of the component, the
FWHM major axis of the circular component, and the position angle of the major axis of the component.} \label{0605modelfits1} {
\begin{tabular}{clllll}
\hline \noalign{\smallskip}
{\small Epoch} & {\small Id.} & {\small S }      & {\small r}     & {\small $\theta$ } & {\small Ma.A. }\\
{\small [yr]} &  & {\small [Jy]}      & {\small [mas]}     & {\small  [deg]} & {\small [mas]}\\
\noalign{\smallskip} \hline \noalign{\smallskip} \hline \noalign{\smallskip}
1995.57   & $D$  & 2.533$\pm$0.382 & 0.0$\pm$0.04& 0.0                & 0.32 \\
          &      & 0.045$\pm$0.015 & 1.35$\pm$0.06  & 103.59$\pm$4.51 & 0.81  \\
          & $C1$ & 0.084$\pm$0.014 & 3.20$\pm$0.02  & 108.79$\pm$0.41 & 0.33  \\
\hline
1996.82   & $D$  & 2.001$\pm$0.319 & 0.0$\pm$0.16  & 0.0              & 0.55  \\
          &      & 0.045$\pm$0.022 & 1.70$\pm$0.07  & 110.22$\pm$2.05 & 0.20  \\
          &      & 0.045$\pm$0.007 & 2.82$\pm$0.04  & 125.70$\pm$0.90 & 0.30  \\
          & $C1$ & 0.067$\pm$0.011 & 3.19$\pm$0.05  & 107.67$\pm$0.97 & 0.53  \\
\hline
1998.83   & $D$  & 0.770$\pm$0.107 & 0.0$\pm$0.03 & 0.0               & 0.10  \\
          & $d2$ & 0.281$\pm$0.035 & 0.50$\pm$0.01  & 123.65$\pm$1.92 & 0.43  \\
          & $d3$ & 0.248$\pm$0.047 & 1.38$\pm$0.05  & 127.61$\pm$0.86 & 0.51  \\
          & $C1$ & 0.099$\pm$0.014 & 3.80$\pm$0.04  & 108.49$\pm$0.86 & 0.76  \\
          & $d6$ & 0.022$\pm$0.009 & 6.43$\pm$0.37  & 112.29$\pm$3.20 & 2.59  \\
          & $d7$ & 0.008$\pm$0.001 & 8.38$\pm$0.43  & 98.75$\pm$3.01  & 0.84  \\
          &      & 0.011$\pm$0.002 & 9.95$\pm$0.98  & 108.37$\pm$5.50 & 1.66  \\
\hline
2001.20   & $D$  & 0.978$\pm$0.144 & 0.0$\pm$0.02 & 0.0               & 0.11  \\
          & $d1$ & 0.384$\pm$0.008 & 0.43$\pm$0.01  & 126.68$\pm$1.09 & 0.39  \\
          & $d2$ & 0.073$\pm$0.028 & 1.69$\pm$0.07  & 130.95$\pm$1.01 & 0.97  \\
          & $d3$ & 0.058$\pm$0.044 & 2.32$\pm$0.08  & 117.61$\pm$1.18 & 1.08  \\
          & $C1$ & 0.013$\pm$0.002 & 3.61$\pm$0.17  & 126.00$\pm$2.70 & 0.61  \\
          &          & 0.109$\pm$0.013 & 4.22$\pm$0.05  & 112.50$\pm$0.76 & 1.34  \\
          & $d6$ & 0.011$\pm$0.002 & 6.71$\pm$0.27  & 117.50$\pm$2.30 & 0.87  \\
          & $d7$ & 0.013$\pm$0.002 & 8.83$\pm$0.29  & 104.45$\pm$1.90 & 1.21  \\
\hline
2003.09   & $D$  & 1.145$\pm$0.167 & 0.0$\pm$0.02& 0.0                & 0.22  \\
          & $d0$ & 0.048$\pm$0.007 & 0.60$\pm$0.02  & 151.77$\pm$1.70 & 0.21  \\
          & $d1$ & 0.150$\pm$0.042 & 1.24$\pm$0.10  & 129.60$\pm$1.99 & 1.01  \\
          & $d2+C1$ & 0.039$\pm$0.007 & 2.95$\pm$0.08  & 123.91$\pm$1.40 & 0.77  \\
          & $d3$ & 0.051$\pm$0.044 & 4.25$\pm$0.08  & 110.50$\pm$2.68 & 2.14  \\
\hline
2004.44   & $D$  & 0.843$\pm$0.119 & 0.0$\pm$0.07& 0.0                & 0.08  \\
          & $da$ & 0.376$\pm$0.067 & 0.30$\pm$0.03  & 133.42$\pm$3.42 & 0.15  \\
          & $d0$ & 0.141$\pm$0.022 & 1.31$\pm$0.06  & 133.02$\pm$2.01 & 1.17  \\
          & $d1$ & 0.031$\pm$0.003 & 1.79$\pm$0.02  & 110.46$\pm$0.63 & 0.10  \\
          & $d2+C1$ & 0.096$\pm$0.014 & 3.72$\pm$0.12  & 118.60$\pm$1.11 & 1.98  \\
          & $d3$ & 0.058$\pm$0.002 & 5.41$\pm$0.09  & 113.42$\pm$0.73 & 1.18  \\
          & $d6$ & 0.009$\pm$0.001 & 7.85$\pm$0.18  & 118.32$\pm$1.30 & 0.56  \\
          & $d7$ & 0.019$\pm$0.010 & 9.42$\pm$0.39  & 108.10$\pm$6.79 & 3.76  \\
\hline
2005.17   & $D$  & 0.835$\pm$0.119 & 0.0$\pm$0.06& 0.0                & 0.05  \\
          & $da$ & 0.323$\pm$0.106 & 0.40$\pm$0.06  & 133.26$\pm$5.41 & 0.08  \\
          & $d0$ & 0.061$\pm$0.009 & 1.20$\pm$0.17  & 134.21$\pm$2.60 & 0.74  \\
          & $d1$ & 0.081$\pm$0.011 & 2.10$\pm$0.07  & 121.01$\pm$1.27 & 1.01  \\
          & $d2+C1$ & 0.100$\pm$0.016 & 4.28$\pm$0.06  & 121.57$\pm$0.87 & 1.50  \\
          & $d3$ & 0.033$\pm$0.005 & 5.75$\pm$0.08  & 114.90$\pm$0.80 & 0.75  \\
          &      & 0.007$\pm$0.002 & 6.18$\pm$0.09  & 106.30$\pm$0.90 & 0.13  \\
          & $d6$ & 0.027$\pm$0.006 & 7.73$\pm$0.33  & 103.00$\pm$2.50 & 2.94  \\
          &      & 0.005$\pm$0.001 & 11.73$\pm$0.34 & 103.70$\pm$1.70 & 0.48  \\
          &      & 0.007$\pm$0.001 & 13.26$\pm$0.31 & 101.00$\pm$1.40 & 0.49  \\
          &      & 0.012$\pm$0.002 & 19.83$\pm$1.24 & 111.90$\pm$3.60 & 3.99  \\
\hline
2005.56   & $D$  & 0.907$\pm$0.132 & 0.0$\pm$0.03& 0.0                & 0.09  \\
          & $da$ & 0.358$\pm$0.058 & 0.44$\pm$0.02  & 135.27$\pm$1.18 & 0.16  \\
          & $d0$ & 0.084$\pm$0.017 & 1.77$\pm$0.05  & 126.35$\pm$1.53 & 1.13 \\
          & $d1$ & 0.019$\pm$0.006 & 2.54$\pm$0.08  & 116.81$\pm$1.83 & 0.42  \\
          & $d2+C1$ & 0.103$\pm$0.021 & 4.32$\pm$0.15  & 121.36$\pm$1.88 & 1.38  \\
          & $d3$ & 0.058$\pm$0.007 & 5.75$\pm$0.13  & 112.87$\pm$0.77 & 1.17  \\
          & $d6$ & 0.009$\pm$0.001 & 8.04$\pm$0.55  & 104.90$\pm$4.00 & 1.68  \\
          & $d7$ & 0.004$\pm$0.001 & 10.41$\pm$1.19 & 95.45$\pm$6.50  & 1.46  \\
          &          & 0.008$\pm$0.001 & 11.12$\pm$0.43 & 110.02$\pm$2.20 & 1.05  \\
          &          & 0.003$\pm$0.001 & 21.24$\pm$3.66 & 113.30$\pm$9.80 & 3.38  \\
\hline
2005.71   & $D$  & 0.822$\pm$0.123 & 0.0$\pm$0.04   & 0.0             & 0.13  \\
          & $da$ & 0.288$\pm$0.044 & 0.44$\pm$0.02  & 132.92$\pm$1.41 & 0.22  \\
          & $d0$ & 0.086$\pm$0.011 & 1.77$\pm$0.04  & 127.87$\pm$1.42 & 1.01  \\
          & $d1$ & 0.026$\pm$0.004 & 2.61$\pm$0.05  & 118.97$\pm$1.20 & 0.43  \\
          & $d2+C1$ & 0.053$\pm$0.008 & 4.28$\pm$0.06  & 124.57$\pm$1.18 & 1.08  \\
          & $d3$ & 0.108$\pm$0.016 & 5.31$\pm$0.09  & 114.02$\pm$0.98 & 2.10  \\
\noalign{\smallskip} \hline

         \end{tabular}
     }
     \end{center}
     \end{table*}

\section{Precession model} \label{precess}
\begin{figure}[tb] \centering  
\includegraphics[clip,width=9.0cm]{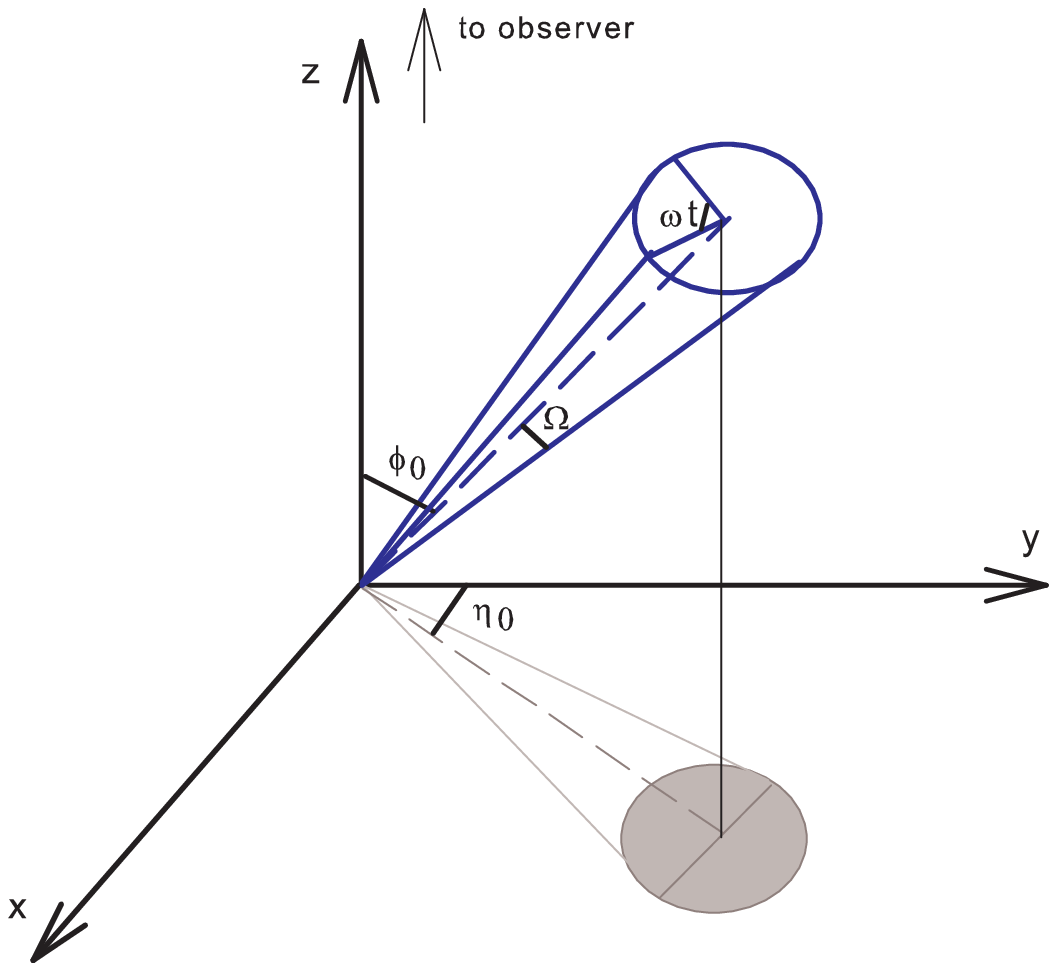}
\caption{Jet precession geometry. The line of sight is parallel to the z-axis.} \label{0605_model_plot}
\end{figure}

The probable periodical variability of the radio total flux-densities of B0605$-$085 and the helical motion of the quasi-stationary jet
component $C1$ can be explained by jet precession. We used a precession model, described in Abraham \& Carrara (1998) and Caproni \& Abraham
(2004) for fitting the helical path of the jet component $C1$. In this model, the rectangular coordinates of a precession cone in the source
frame are changing with time $t'$ due to precession:

\begin{eqnarray}
X(t') = [\cos \Omega \sin \phi_{0} + \sin \Omega \cos \phi_{0} \sin \omega t'] \cos \eta_{0} - \nonumber\\
\sin \Omega \cos \omega t' \sin \eta_{0}, \label{eq:x_t}
\end{eqnarray}

\begin{eqnarray}
Y(t') = [\cos \Omega \sin \phi_{0} + \sin \Omega \cos \phi_{0} \sin \omega t'] \sin \eta_{0} + \nonumber\\
\sin \Omega \cos \omega t' \cos \eta_{0}, \label{eq:y_t}
\end{eqnarray}
where $\Omega$ is the semi-aperture angle of the precession cone, $\phi_{0}$ is the angle between the precession cone axis and the line of
sight and it is actually the average viewing angle of the source, and $\eta_{0}$ is the projected angle of the cone axis onto the plane of the
sky (see Fig.~\ref{0605_model_plot}). The angular velocity of the precession is $\omega = 1 / P$, where $P$ is the period. We take a $P$ value
of 7.9-year (timescale found in the total flux-density radio light curves and in the helical movement of $C1$). The time in the source frame
($t'$) and the frame of the observer ($t$) are related by the Doppler factor $\delta$ as

\begin{equation}
\Delta t' = \frac{\delta(\phi,\gamma)}{1+z} \Delta t. \label{eq:time}
\end{equation}

\begin{figure*}[htb] \centering  
\hbox{
\includegraphics[clip,width=9.0cm]{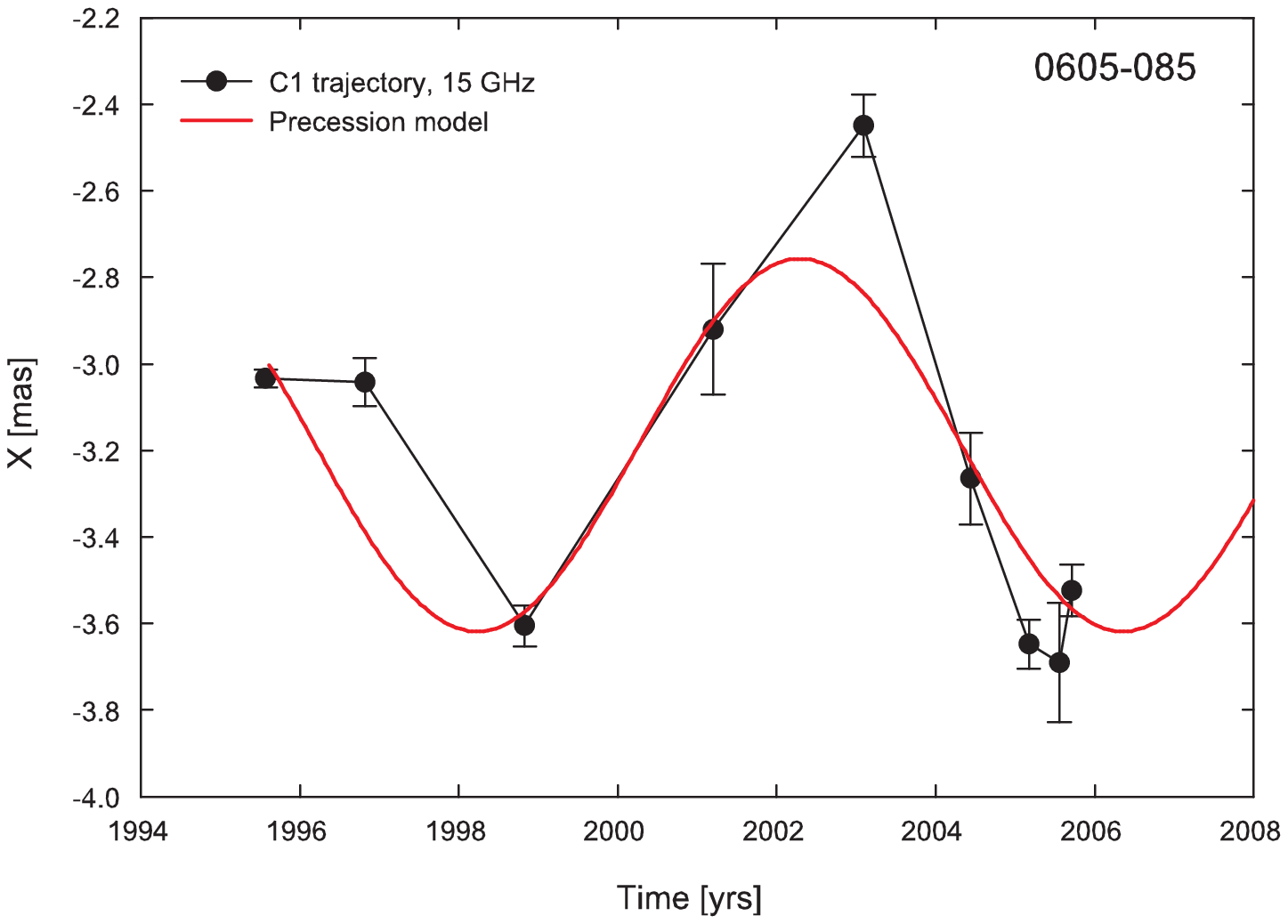}
\includegraphics[clip,width=9.0cm]{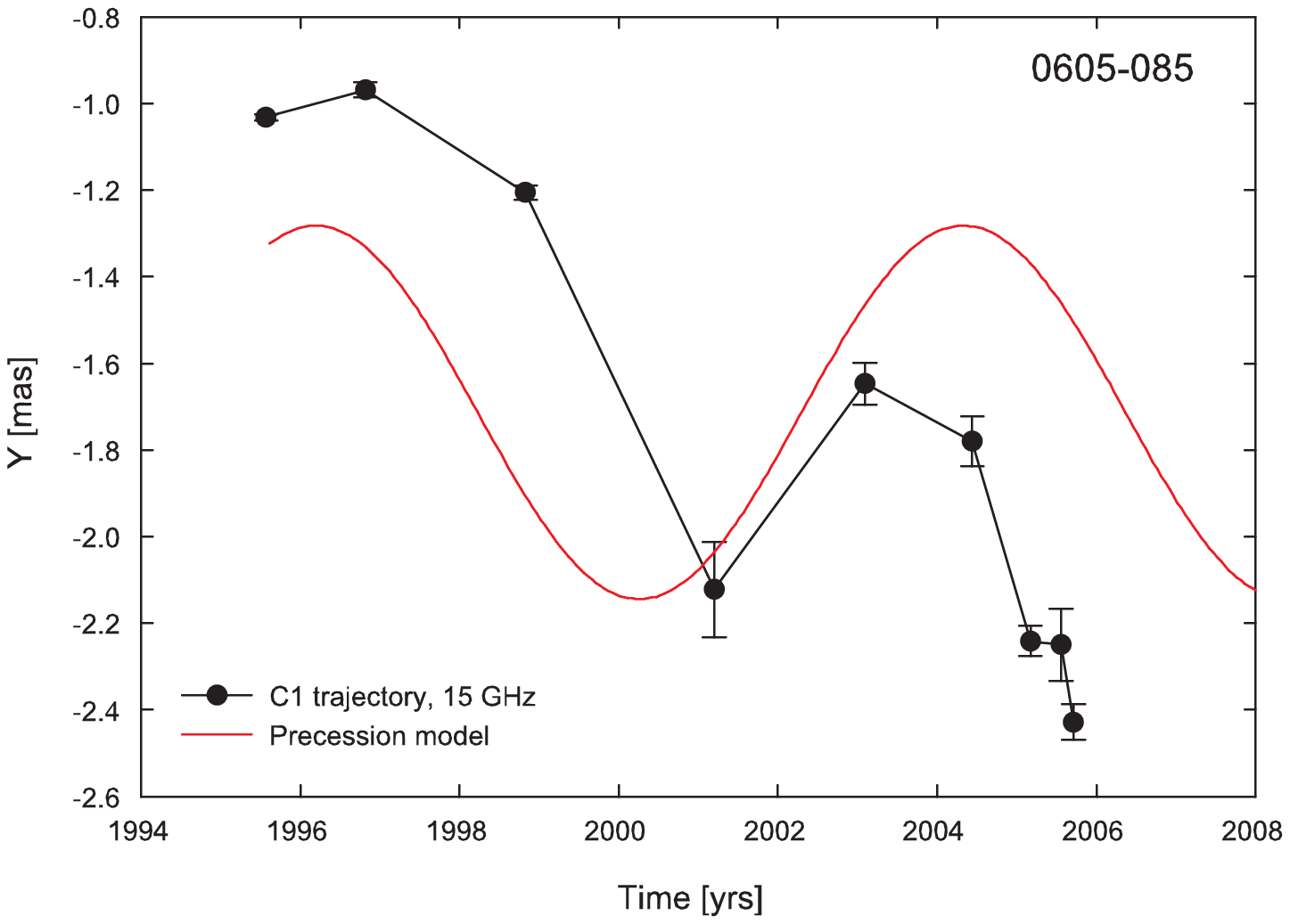}
}\caption{Sky trajectory of the quasi-stationary component $C1$ in right ascension (Left) and declination (Right). The solid line shows the fit
of the precession model. The precession model fits well the trajectory of a quasi-stationary jet component $C1$} \label{0605_precession_model}
\end{figure*}

However, the only time-dependent term in the equations~(\ref{eq:x_t}) and~(\ref{eq:y_t}) is $\omega t$, which does not depend on this time
corrections since $\omega \sim 1/t$. Therefore, we can fit the trajectory in the rectangular coordinate $X(t)$ of the quasi-stationary jet
component $C1$, using the formula \ref{eq:x_t}. We assume that the motion of $C1$ reflects the movement of the jet. The non-linear least
squares method was used for the precession model fitting. The angular velocity of the precession $\omega$ and the redshift of the source $z$
are known and we fix them during the fit, whereas the aperture angle of the precessing cone $\Omega$, $\phi_{0}$, and $\eta_{0}$ are used as
free parameters. The results of the fit are shown in Fig.~\ref{0605_precession_model} (Left) and the parameters of the precession model for the
parsec jet of B0605$-$085 are listed in Table~\ref{0605_precession}. The precession model fits well the trajectory of the quasi-stationary jet
component $C1$. The same parameters were used to describe the motion of $C1$ in declination. The model for the declination is shown in
Fig.~\ref{0605_precession_model} (Right).

We found another set of parameters $\Omega = 92.2^\circ$, $\phi_{0} = 69.8^\circ$, $\eta_{0} = 12.9^\circ$ as one of the possible solutions.
However, the viewing angle is very large. Using the equation
\begin{equation}
\beta_{app} = \frac{\beta \sin\phi}{1 - \beta \cos\phi}, \label{eq:speed}
\end{equation}
and taking average value of apparent speeds $\beta_{app,aver} = 16.5c$, we can calculate the speed in the source frame $\beta = 2.49c$, which
is higher then the speed of light. Precession cone aperture angle $\Omega = 92.2^\circ$ is also unlikely. Therefore, this set of parameters is
not plausible. The solution in Table~\ref{0605_precession} is the only physically feasible, since all the parameter space was covered during
the fitting procedure.

\begin{table}[htb]
\begin{center}
\caption{Precession model parameters (see Sect.~\ref{precess}).} \label{0605_precession}
\medskip
\begin{tabular}{lll}
\hline \noalign{\smallskip}
Parameter & Description & Value \\
\hline \noalign{\smallskip}
P & Precession period, fixed & 7.9 yr \\
$\Omega$ & Aperture angle, variable & $23.9^\circ \pm 1.9^\circ$ \\
$\phi_{0}$ & Viewing angle, variable & $2.6^\circ \pm 2.2^\circ$\\
$\eta_{0}$ & Projection angle, variable & $33.6^\circ \pm 6.5^\circ$ \\
\hline
\end{tabular}
\end{center}
\end{table}

\section{Discussion}

Assuming that the helical motion of the quasi-stationary jet component $C1$ is caused solely by jet precession (with a period of 8 years), we
were able to determine the geometrical parameters of the jet precession, such as the aperture angle of a precession cone, the viewing angle,
and the projection angle. The fitted viewing angle $\phi_{0} = 2.6^\circ \pm 2.2^\circ$ agrees well with the upper limit for the viewing angle
deduced from the apparent speeds of the jet components $\phi_\mathrm{max} = 3.7^\circ$. It is also similar to the viewing angle
$\phi_\mathrm{var} = 5^\circ$ derived from the radio total flux-density variability by Hovatta et al. (2008). The jet precession will also
cause the periodical variability of the radio total flux-density light curves with the same period of 8 years, due to periodical changes of the
Doppler factor. The flux density $S_{j}$ is changing with the Doppler factor due to the jet precession $S_{j} = S_{j}' \delta (\phi,
\gamma)^{p+\alpha}$, where $\alpha$ is the spectral index (Lind \& Blandford 1985). From this equation we can expect that the flares due to the
jet precession will appear simultaneously at different frequencies, which is actually observed for the major outbursts in 1973, 1988, and
1995-1996. Therefore, the jet precession is probably responsible for the total flux-density variability, periodicity in the outbursts, and the
helical motion of the quasi-stationary jet component $C1$.

Some questions remain unclear. It is still not known why the four major outbursts in the total flux-density light curve which repeat
periodically with an 8-year period have different spectral properties. Longer monitoring of this quasar at several frequencies is necessary in
order to understand spectral properties of the flares.

The precession of B0605$-$085 jet can be caused by several physical mechanisms. Jet instabilities, a secondary black holes rotating around the
primary black hole, and accretion disc instabilities might be responsible for the observed changes of jet direction (e.g., Camenzind \&
Krockenberger 1992, Hardee \& Norman 1988, Lobanov \& Roland 2005). At the moment it is impossible to claim exactly which mechanism can produce
the observed jet precession. It is worth to notice, however, that the possible double-peak structure of the flares at 14.5~GHz in B0605$-$085
is similar to the double-peak flares of OJ~287 observed in optical light curves which were explained by the orbiting secondary black hole. At
the moment we can not say whether there is a secondary black hole in the center of a hosting galaxy in B0605$-$085 and further observations and
theoretical modelling are needed.

The next total flux-density flare of the source might appear in 2012 if the 8-year period will preserve over time. The 8-year period in the
flux-density variability has predicted the next powerful outburst to appear in $\sim$ 2004. However, the flare was of a very low flux level
which might be due to long-term trend in flux-density variability. Recent observations of B0605$-$085 have shown a rise in fluxes which might
suggest a probable peak close to 2012 taking into account the duration of the outbursts in this source. This will mean that the 8-year period
interferes with long-term changes of the flux-densities. Further radio total flux-density and VLBI observations are needed in order to check
this prediction.

\section{Summary}
Our analysis of the jet kinematics in B0605$-$085 shows:
\begin{itemize}
\item We find strong evidence for a period of the order of 8 years ($7.9\pm0.5$ years when averaged over frequencies) in the total flux-density
light curves at 4.8~GHz, 8~GHz and 14.5~GHz, which was observed for four cycles.

\item The measurement of frequency-dependent time delays for the brightest outbursts in 1988, and 1995-1996 and the visual analysis for the
1973 flare shows that these flares appeared almost simultaneously at all frequencies.

\item The spectral analysis of the flares shows that the main flares in 1973, 1988, 1995-1996 have a flat spectrum, whereas the flare in 1981
has a steep spectrum, what suggests that the four cycles of the 8-year period have different spectral properties.

\item The average instantaneous speeds of the jet components reveal a helical pattern along the jet axis with a characteristic scale of 3~mas.
This scale corresponds to a time scale of about 7.7 years, which is similar to the periodicity timescale found in the total flux-densities.

\item The quasi-stationary jet component $C1$ follows a helical path with a period of 8 years. This time scale coincides with the period found
in the total flux-density light curves.

\item We found evidence that the jet components of B0605$-$085 follow two different types of trajectories. Some components move along the
straight lines, whereas other components follow significantly curved paths.

\item The fit of the precession model to the trajectory of the quasi-stationary jet component $C1$ has shown that it can be explained by
changes of the jet direction with a precession period of $P = 8$ yrs (in the observers frame), aperture angle of the precession cone $\Omega =
23.9^\circ \pm 1.9^\circ$, viewing angle $\phi_{0} = 2.6^\circ \pm 2.2^\circ$, and projection angle $\eta_{0} = 33.6^\circ \pm 6.5^\circ$.

\end{itemize}

\begin{acknowledgements}
This research has made use of data from the MOJAVE database that is maintained by the MOJAVE team (Lister et al., 2009, AJ, 137, 3718). N. A.
Kudryavtseva and M. Karouzos were supported for this research through a stipend from the International Max Planck Research School (IMPRS) for
Astronomy and Astrophysics. We would like to thank Christian Fromm for fruitful discussions and help with the picture. This research has made
use of data from the University of Michigan Radio Astronomy Observatory which is supported by the National Science Foundation and by funds from
the University of Michigan. The Alonquin Radio Observatory is operated by the National Research Council of Canada as a national radio astronomy
facility. The operation of the Haystack Observatory is supported by the grant from the National Science Foundation. The National Radio
Astronomy Observatory is a facility of the National Science Foundation operated under cooperative agreement by Associated Universities, Inc.
This publication has emanated from research conducted with the financial support of Science Foundation Ireland. This research has made use of
the SIMBAD database, operated at CDS, Strasbourg, France. We thank the anonymous referee for a careful reading of the manuscript and useful
comments and suggestions that have improved this paper.
\end{acknowledgements}

\end{document}